\documentclass[twocolumn,floatfix,amsfonts,amssymb,notitlepage]{revtex4-1}
\usepackage{color}
\usepackage{graphicx}
\usepackage{amsmath}
\usepackage{latexsym}
\usepackage{epstopdf}
\usepackage{amsmath}
\usepackage{amssymb}
\usepackage{dsfont}
\usepackage{multirow}
\usepackage{mathtools}
\usepackage{tcolorbox}
\usepackage{cleveref}
\usepackage{ulem}

\newcommand{\beq}{\begin{equation}}
\newcommand{\eeq}{\end{equation}}
\newcommand{\bea}{\begin{eqnarray}}
\newcommand{\eea}{\end{eqnarray}}

\def\ket#1{|#1\rangle}

\newcommand{\bebox}{\begin{tcolorbox}}
\newcommand{\eebox}{\end{tcolorbox}}

\newcommand{\eq}{\begin{equation}}
\newcommand{\en}{\end{equation}}
\newcommand{\ear}{\begin{eqnarray}}
\newcommand{\rae}{\end{eqnarray}}

\newcommand{\rf}[1]{(\ref{#1})}

\def\ket#1{|#1\rangle}

\newcommand{\be}{\begin{eqnarray}}
\newcommand{\ee}{\end{eqnarray}}

\makeindex

\begin{document}
\title{Free fermionic and parafermionic multispin quantum chains with non-homogeneous interacting ranges}
\author{Francisco C. Alcaraz}
\email{alcaraz@ifsc.usp.br}
\affiliation{ Instituto de F\'{\i}sica de S\~{a}o Carlos, Universidade de S\~{a}o Paulo,
Caixa Postal 369, 13560-970, S\~{a}o Carlos, SP, Brazil}
\date{\today{}}

\begin{abstract}
  A large family of multispin interacting one-dimensional quantum spin models 
with $Z(N)$ symmetry and a free-particle eigenspectra are known in the 
literature. They are free-fermionic ($N=2$) and free-parafermionic ($N\geq 2$) 
quantum chains. The essential ingredient that implies the free-particle 
spectra is the fact that these Hamiltonians are expressed in terms of 
generators of a $Z(N)$ exchange algebra. In all these known quantum chains 
the number of spins in all the multispin interactions (range of 
interactions) is the same and therefore,
the models have homogeneous interacting range. 
In this paper we extend the $Z(N)$ exchange algebra, by introducing new models 
with a free-particle spectra, where the interaction ranges of the 
multispin interactions are not uniform anymore and depends on the lattice sites 
(non-homogeneous interacting range). We obtain the general conditions that 
the site-dependent ranges of the multispin interactions have to satisfy to 
ensure a free-particle spectra. Several simple examples are introduced.  
We study in detail the critical properties in the case where the range of 
interactions of the even (odd) sites are constant. The dynamical critical 
exponent is evaluated in several cases.

\end{abstract}

\maketitle

\section{Introduction} 
In condensed matter physics and statistical mechanics there exist a class 
of interesting models, that besides being exact integrable, have a 
free-particle eingenspectra. They are considered as free models because 
their eigenspectra are given by combinations of independent 
pseudo-energies. In the most known cases the models are solved by the standard 
Jordan-Wigner transformation \cite{lieb,pfeuty}. This transformation allow 
us to produce an effective model, given by the addition of bilinear 
fermionic operators, whose eigenspectra solution follows from a generalized 
Fourier transform (Bogoliubov transformation). However, more recently 
a larger class of free-particle quantum chains was introduced. They do not 
have anymore a bilinear fermionic  form, after a Jordan-Wigner transformation. These 
are quantum spin models defined in terms of $Z(N)$ fermionic ($N=2$) or parafermionic operators 
($N=3,4,...$), with  multispin interacting couplings. For the cases 
where $N=2$ the models are Hermitian, but for $N>2$ they are in general 
non-Hermitian. 
 The exact solution of these models is known only in the case of open 
boundary conditions (OBC). These models are known  to have a 
free parafermionic  eigenspectra when all the 
 multispin interactions, whose coupling constants 
may depend on the lattice sites, 
couple always the same number of spins.  The number of spin in the multispin interaction is ($p+1$)  ($p=1,2,\ldots$). In the case $p=1$ and 
$N=2$ the models recover the free fermionic quantum chains with two spin 
interactions, like the quantum Ising chain in a transverse field \cite{pfeuty}. The case 
$p=2$ and $N=2$ give us the three-spin interaction Fendley model 
\cite{fendley2}, also known
as a disguise fermion model. The cases $p=1$ and $N>2$ are the 
free parafermionic Baxter models \cite{baxter1,baxter2, fendley1,baxter3,perk1,perk2,AB1,AB2}. The general cases where $p$ and $N$ are arbitrary was solved 
in \cite{AP1,AP2} (see also \cite{AP3,ising-analogues}), by extending the fermionic case $N=2$ and $p=2$ solved 
by Fendley \cite{fendley2}. Actually these models  with general 
values of $p$ can be considered as particular cases of models defined in 
frustration graphs \cite{network,circuits-pozsgay}. A more general related free-fermion model was also introduced recently in  
\cite{fendley-pozsgay} and \cite{fukai}.
%

It is interesting to mention that although all the above models have a 
free-particle spectra for OBC, the eigenenergies  
are not known in a direct 
form in the cases where the quantum chains are defined on  periodic lattices.  

In this paper we are going to consider general multispin interacting one-dimensional 
 models where the range of interactions (the number of spins coupled in a 
multispin interaction) depend on the 
particular site. The above mentioned ($p+1$) interacting models are particular 
cases where the interaction are site independent (homogeneous). By deriving 
the sufficient conditions for having a free-particle eigenspectra we end up 
with the restrictions the range of interactions have to satisfy to ensure 
a free-particle eigenspectra for the models. We obtain general 
non-homogeneous 
interacting models and  in several examples  their critical behavior are 
derived.

The paper is organized as follows. In section II we introduce the 
one-dimensional models with general interacting range, and show the constraints the ranges have to satisfy in order to produce a free-particle eigenspectra. 
In section III we present several examples of non-homogeneous models and 
derive their critical behavior.  In section IV we draw our conclusions. 
Finally in  apendix A  we  derive the inverse relations for the models. 
A relation that ensures, for the models we considered,  the free-particle eigenspectra.


\section{General non-homogeneous free-particle one-dimensional quantum chains}

\subsection{Multispin models with homogeneous interacting ranges}

In \cite{AP1,AP2} a large family of homogeneous quantum chains was introduced 
with an effective free-particle spectra. These are special models with 
$Z(N)$ symmetry, whose spectra are formed by the composition of fermionic ($N=2$) or 
parafermionic ($N>2$) one-particle pseudo-energies. 

In the above mentioned free-particle models the range of interactions of a 
given spin is uniform (site independent). In this section we search for 
possible extensions, by considering quantum spin chains with multispin interactions where 
the number of spin involved in the interactions (range of interaction) is 
site dependent, and we call such models as inhomogeneous range models. 

The  Hamiltonians  are described by $M$ 
generators $h_i^{(r_i)}$, 
attached to the sites $i=1,2,\ldots,M$, and obeying a $Z(N)$  exchange algebra 
specified by the set of non-negative integers 
$\{r_i\}$ = ($r_1,r_2,\ldots,r_M$). The generators play the 
role of energy density operators and the Hamiltonians are:
\be \label{2.1}
H_M^{(N,\{r_i\})} (\lambda_1,\ldots,\lambda_M) = -\sum_{i=1}^M h_i^{(r_i)}.
\ee 
The scale and range of interactions are encoded in the $Z(N)$ exchange algebra 
satisfied by the generators $\{h_i^{(r_i)}\}$. The integer parameters $r_i>0$ give the 
range of the multispin interactions at the right of the site $i$, of the associative 
algebra:
\be \label{2.2a}
h_i^{(r_i)} h_j^{(r_j)} = \begin{cases} 
{\omega} h_j^{(r_j)} h_i^{(r_i)},&0<(j-i)\leq r_i, \\
 \frac{1}{\omega} h_j^{(r_j)} h_i^{(r_i)}, &0<(i-j)\leq r_j, \\
h_j^{(r_j)} h_i^{(r_i)}, & \mbox{otherwise}. 
\end{cases}
\ee
with $\omega = e^{i2\pi/N}$. The coupling constants in \rf{2.1} 
$\lambda_1,\lambda_2, \ldots,\lambda_M$ are fixed by the closure relation of 
the $Z(N)$ algebra \rf{2.2a}:
\be \label{2.2b}
[h_i^{(r_i)}]^N = \lambda_i^N.
\ee
The standard exchange algebra is the one given by \rf{2.2a} with 
$r_1=r_2=\cdots=r_M=1$.
Notice that actually $r_M$ is arbitrary since we have no generators at the 
right of the site $M$. In the cases where $N=2$ we have free-fermionic 
quantum chains and for $N>2$  free parafermionic ones.

The cases in \rf{2.1} where the ranges $\{r_i\}$ in \rf{2.2a} are 
site independent are known to produce  Hamiltonians with a 
free-particle spectra. Examples of representations of \rf{2.2a}-\rf{2.2b} 
with $r_1=r_2,\ldots=r_M=1$ are given for $N=2$
\bea \label{2.3}
h_{2i-1}^{(1)} &=& \lambda_{2i-1}\sigma_i^x, \quad 
h_{2i}^{(1)} = \lambda_{2i}\sigma_i^z \sigma_{i+1}^z, \nonumber \\
&&[h_i^{(1)}]^2 = \lambda_i^2, \quad i=1,2,\ldots,
\eea
and for arbitrary $N$
\bea \label{2.4}
h_{2i-1}^{(1)} &=& \lambda_{2i-1}X_i, \quad 
h_{2i}^{(1)} = \lambda_{2i}Z_i Z_{i+1}^+, \nonumber \\
&&[h_i^{(1)}]^N = \lambda_i^N, \quad i=1,2,\ldots,
\eea
where $\sigma^x,\sigma^z$ are spin-1/2 Pauli matrices, and $X,Z$ their 
$Z(N)$ generalization satisfying
\bea \label{2.5}
&&X_iZ_i = \omega \; Z_iX_i, \quad [X_i,X_j]=[Z_i,Z_j]=0, \nonumber \\ 
&&X_i^N = Z_i^N=1, \quad Z^+ = Z^{N-1}.
\eea
The representations \rf{2.3} and \rf{2.4} give us, for $M$ odd, the 
standard quantum Ising chain \cite{pfeuty}

\be \label{2.6} 
H^{(2;1,1\ldots,1)}= - \sum_{i=1}^{\frac{M+1}{2}} \lambda_{2i-1} \sigma_i^x 
-\sum_{i=1}^{\frac{M-1}{2}} \lambda_{2i}\sigma_i^z\sigma_{i+1}^z,
\ee
and the $Z(N)$ free parafermionic Baxter quantum chain 
\cite{baxter1,fendley1,baxter2}
\be \label{2.7} 
H^{(2;1,1\ldots,1)}= - \sum_{i=1}^{\frac{M+1}{2}} \lambda_{2i-1} X_i 
-\sum_{i=1}^{\frac{M-1}{2}} \lambda_{2i}Z_i Z_{i+1}^+.
\ee
A representation for the case $r_1=r_2=\cdots=r_M=2$ and $N=2$ give us the 
3-spin interaction Fendley quantum chain \cite{fendley2}, and for general 
$N$ and $r_1=r_2=\cdots=r_M=p$ ($p \geq 2$) we have the representation 
\be \label{2.8}
h_i^{(p)} = \lambda_i \left(\prod_{j=i-p}^{i-1} Z_j\right)X_i, 
\quad Z_{-\ell} =1 \quad (\ell \leq 0),
\ee
that gives the $M$-lattice size free parafermionic Hamiltonian \cite{AP1,AP2}:
\be \label{2.9}
H^{(N;p,p,\ldots,p)} = -\sum_{i=1}^M \lambda_i \left(\prod_{j=i-p}^{i-1} Z_j\right)X_i.
\ee

\subsection{Multispin models with non-homogeneous interacting ranges}

We are going to introduce here the possible generalizations of the 
Hamiltonians \rf{2.1} where the ranges $\{r_i\}$, defining the algebra 
\rf{2.2a}-\rf{2.2b}, are site dependent, giving us non-homogeneous interacting 
range Hamiltonians.

\subsubsection{The constraints obtained from the involution of the 
conserved charges}

The general conditions for the existence of arbitrary free-fermionic models 
were derived in Ref. \cite{network} and \cite{ref-inv-zn}, respectively 
(see also \cite{circuits-pozsgay}). These derivations are obtained by using 
graph theory. 

In this paper we are going to present a {\it direct derivation} of the 
sufficient conditions ensuring a free-particle spectra. We   do not 
need the knowledge of graph theory, but only the simple notion of 
commutators.

Our derivation for the constraints of the algebra are general and not 
restricted to one-dimensional models. For this reason we do not specify 
in the generators their range of interaction, and instead of writing 
$h_i^{(r_i)}$ we write simply $h_i$.

As in the uniform interacting cases $r_1=r_2=\cdots=r_M$, we introduce the algebraic 
``words'' formed 
by the special products of $\ell$ generators (``letters"):
\be \label{2.10}
W^{(\ell)}(i_1,i_2,\ldots,i_{\ell}) = h_{i_1} h_{i_2} \cdots h_{i_{\ell}},
\ee
where all the operators $h_{i_1},h_{i_2}, \ldots,h_{i_{\ell}}$ commute 
among themselves.

We now define the $\ell$-charges ($\ell=0,1,2,\ldots$):
\be \label{2.11}
Q^{(\ell)} = \sum_{i_1,i_2,\ldots,i_{\ell}}^{(*)} W^{(\ell)} (i_1,i_2,\ldots,i_{\ell}),
\ee
where $(*)$ denotes the sum over all independent possibilities of the 
$\ell$-letters words $W^{(\ell)}$. The charge $Q^{(0)}=1$ is the identity operator, and $Q^{(1)} = -H$. 

We now search for all the  constraints  the algebra of $\{h_i^{(r_i)}\}$ 
have to satisfy in order to give  the involution
\bea \label{2.12}
&& [Q^{(\ell)}, Q^{(\ell')}] =  
\sum_{i_1,\ldots,i_{\ell}}^{(*)} 
\sum_{j_1,\ldots,j_{\ell'}}^{(*)} \nonumber \\
&&[W^{(\ell)}(i_1,\ldots,i_{\ell}) , 
W^{(\ell')}(j_1,\ldots,j_{\ell'})] = 0,
\eea	
for $\ell,\ell' \geq 0$.

By a direct calculation of the commutators 
$[W^{(\ell)},W^{(\ell')}]$ we will search for the general conditions 
required for the algebra of $h_i^{(r_i)}$ that ensure its involution \rf{2.12}. 

We consider separately the several types of commutations in \rf{2.12}:
\bea \label{2.13}
C(i_1,\ldots,i_{\ell}&;&j_1,\ldots,j_{\ell'}) = \nonumber \\ 
&&[W^{(\ell)}(i_1,\ldots,i_{\ell}), W^{(\ell')}(j_1,\ldots,j_{\ell'})].
\eea

{\bf{a)}} 
If $\{h_{i_1},\ldots,h_{i_{\ell}}\}$ commute with 
$\{h_{j_1},\ldots,h_{j_{\ell'}}\}$, then
\be \label{2.14}
C(i_1,\ldots,i_{\ell};j_1,\ldots,j_{\ell'}) =0,
\ee
and we have no constraints.

{\bf{b)}} 
If in the set $\{h_{j_1},\ldots,h_{j_{\ell'}}\}$ there exists a single 
operator $h_{j_k}$ that does not commute with the single operator 
$h_{i_m}$ in the set $\{h_{i_1},\ldots,h_{i_{\ell}}\}$, we may consider in the 
sum \rf{2.12} the combinations
\bea \label{2.15a}
C(i_1,\ldots,i_{\ell};&& j_1,\ldots,j_{\ell'}) \nonumber  \\
&&+ 
C(i_1,\ldots,i_{\ell};j_1,\ldots,j_{\ell'})|_{h_{j_k} \leftrightarrow h_{i_m}} ,
\eea
where we used the notation 
\bea \label{2.15}
C(A,B,C,\ldots &;&F,G,H,\ldots)|_{B\leftrightarrow H} = \nonumber \\
&&C(A,H,C,\ldots;F,G,B,\ldots).
\eea

The combination \rf{2.15a} vanishes since it is proportional to 
\be \label{2.17}
[h_{j_k},h_{i_m}] +
[h_{i_m},h_{j_k}] =0.
\ee

\begin{figure} [htb]
\centering
\includegraphics[width=0.45\textwidth]{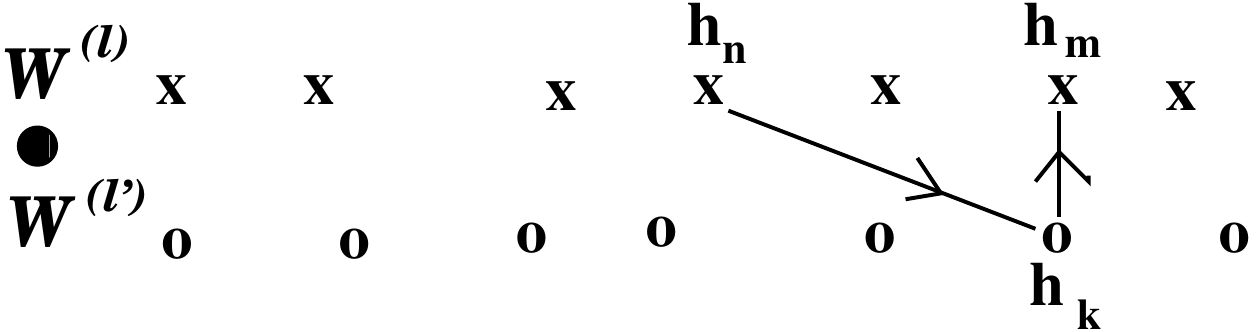}
\caption{
Representations of the product $W^{(\ell)}\cdot W^{(\ell')}$. The crosses and the circles are the generators in $W^{(\ell)}$ and $W^{(\ell')}$, respectively. The 
links connect the generators that do not commute. The arrows  give the 
directions of the multiplication rule in the algebra \rf{2.19}.} \label{Q1}
\end{figure}

{\bf{c)}}
If there is a single operator $h_k$ in $W^{(\ell')}(j_1,\ldots,j_{\ell'})$ that 
 does not commute with two operators ($h_n,h_m$) in
$W^{(\ell)}(i_1,\ldots,i_{\ell})$, then 
$C(i_1,\ldots,i_{\ell};j_1,\ldots,j_{\ell'})$ vanishes only if 
\be \label{2.18}
[h_n\;h_m,h_k] =0.
\ee
This condition is satisfied, for example, if the operators satisfy the 
exchange algebra 
\be \label{2.19}
h_nh_k = \rho h_kh_n \quad \mbox{and} \quad h_mh_k=\rho^{-1} h_k h_m,
\ee
with $\rho$ being an arbitrary complex number (do not need to be a complex 
phase!).
In Fig.~\ref{Q1} we represent the product 
$W^{(\ell)}(i_1,\ldots,i_{\ell})\cdot W^{(\ell')}(j_1,\ldots,j_{\ell'})$. 
The crosses (circles) are the generators in $W^{(\ell)}$ ($W^{(\ell')}$). 
The arrows on the links give the directions for the multiplications in 
\rf{2.19}. 

{\bf{d)}}
If there is a single generator $h_k$ in 
$W^{(\ell')}(j_1,\ldots,j_{\ell'})$ that does not commute with three generators 
$h_m,h_n,h_o$ in $W^{(\ell)}(i_1,\ldots,i_{\ell})$, the involution 
\rf{2.12} imply the constraint 
\be \label{2.20}
[h_mh_nh_o,h_k]=h_m[h_nh_o,h_k]+[h_m,h_k]h_nh_o =0.
\ee
The condition \rf{2.18}, in case c), for the words containing $h_n,h_o$ 
in $W^{(\ell)}$ and $h_k$ in $W^{(\ell')}$ already give us 
$[h_nh_o,h_k]=0$. However, since $[h_m,h_k]\neq 0$ the condition \rf{2.20} is 
{\it not satisfied}. 

\begin{figure} [htb]
\centering
\includegraphics[width=0.45\textwidth]{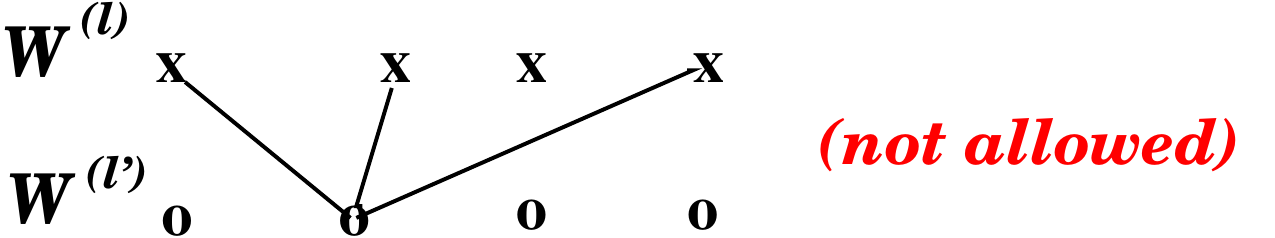} 
\caption{
Representation of the product $W^{(\ell)}\cdot W^{(\ell')}$. The links connect 
generators that do not commute. This configuration  is not allowed, 
since there is three commuting operators in $W^{(\ell)}$ (crosses) that 
do not commute with a single operator in $W^{(\ell')}$ (circles).} \label{Q2}
\end{figure}

This means that {\it{we should not have a single generator $h_i$ that do not 
commute with three other commuting ones}}. In Fig.~\ref{Q2} we illustrate this 
condition. 

\begin{figure} [htb] 
\centering
\includegraphics[width=0.45\textwidth]{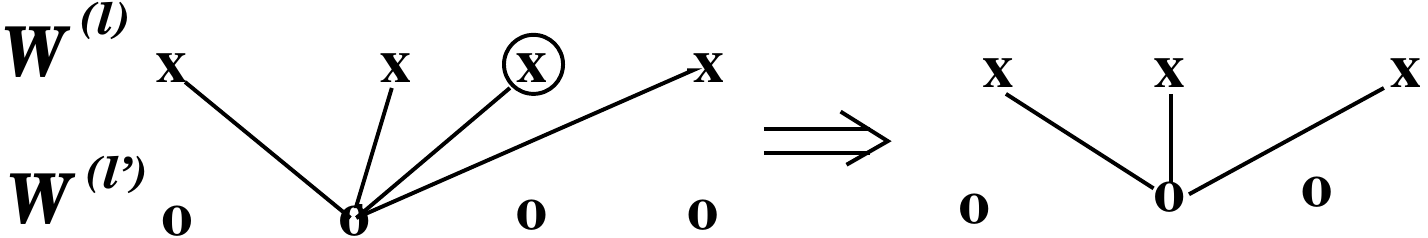}
\caption{
The existence of a word $W^{(\ell)}$ with four commuting operators that does 
not commute with a single operator imply  the existence of a related 
word (see the right of the figure) with three commuting operators, that is  forbidden due to the condition d) 
(see Fig.~\ref{Q2}). } \label{Q3}
\end{figure}

This last condition also implies that we cannot have a single generator 
in $h^{(\ell')}$ that do not commute with three or more  commuting generators 
in $W^{(\ell)}$. In such cases there exist other related $W^{(\ell)}$ 
and $W^{(\ell')}$ words that a single operator in $W^{(\ell')}$ will not 
commute with three others in $W^{(\ell)}$. In Fig.~\ref{Q3} we illustrate 
this.

\begin{figure} [htb]
\centering
\includegraphics[width=0.45\textwidth]{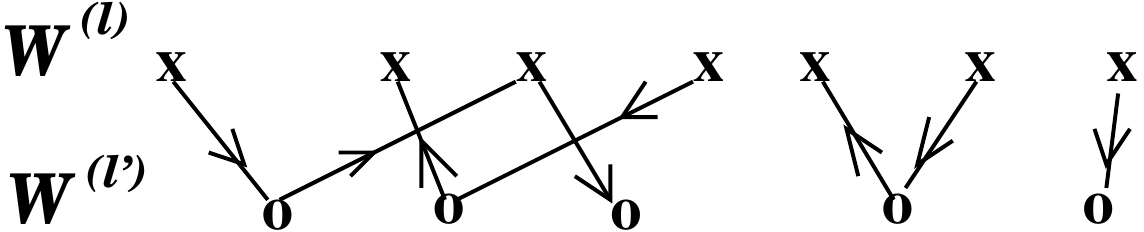} 
\caption{
The clusters of non-commuting operators in the product 
$W^{(\ell)}\cdot W^{(\ell')}$. In the figure we have three clusters with 
($n_c=4,n_{c'}=3$), ($n_c=2,n_{c'}=1$) and ($n_c=1,n_{c'}=1$) 
generators in $W^{(\ell)}$ and $W^{(\ell')}$.} \label{Q4} 
\end{figure}

In the process of commuting $[W^{(\ell)},W^{(\ell')}]$ we have clusters 
formed by the generators in $W^{(\ell)}$ and $W^{(\ell')}$ that do not
commute. 
A given cluster is formed by $n_c$ generators in $W^{(\ell)}$ and $n_{c'}$ generators in $W^{(\ell')}$. 
In Fig.~\ref{Q4}  we have three clusters.
 The clusters have ($n_c=4,n_{c'}=3$), ($n_c=2,n_{c'}=1$) and 
($n_c=1, n_{c'}=1$) generators in $W^{(\ell)}$ and $W^{(\ell')}$.
  As before 
the arrows denote the multiplication directions of the 
exchange algebra \rf{2.19}.

\begin{figure} [htb]
\centering
\includegraphics[width=0.45\textwidth]{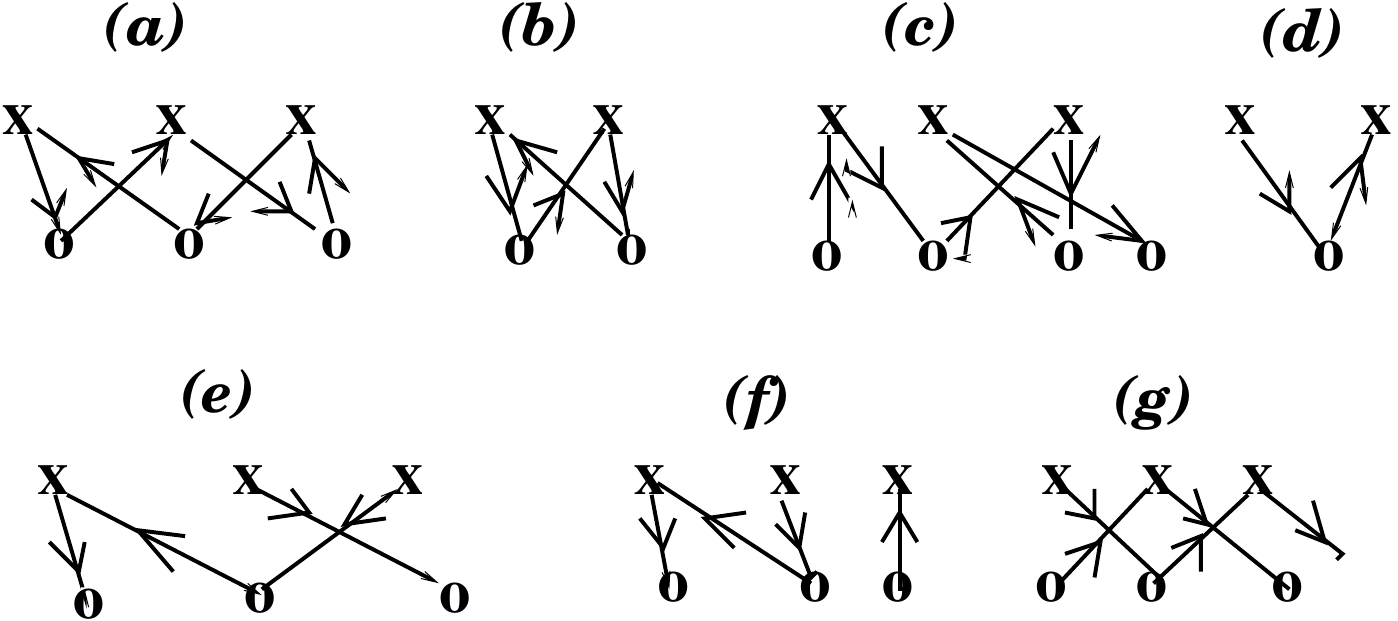}
\caption{
Examples of clusters formed in product $W^{(\ell)}\cdot W^{(\ell')}$. The clusters 
in (a) and (b) are connected even clusters and  (c) and (d) are connected 
odd clusters. The clusters (e), (f) and (g) are unconnected ones. }\label{Q5}
\end{figure}

We define by "connected clusters"  the ones where we visit all their non 
commuting generators in $W^{(\ell)}$ and $W^{(\ell')}$ by following the 
directions of the arrows defining the exchange algebra. The clusters 
$a,b,c,d$ in Fig.~\ref{Q5} are examples of connected ones. 

It is not difficult to convince ourselves that the constraint \rf{2.20} 
imply that any connected cluster formed by $n_c$ and $n_{c'}$ generators in $W^{(\ell)}$ and $W^{(\ell')}$ should obey the restriction $n_c -n_{c'} = -1,0,1$.

Suppose in the product $W^{(\ell)}\cdot W^{(\ell')}$ we have a single 
connected cluster with equal number $n_c=n_{c'}$ of generators in $W^{(\ell)}$ 
and $W^{(\ell')}$, as in cluster ($a$) and ($b$) of Fig.~\ref{Q5}. 
We call them connected even clusters. 
Due to the restriction \rf{2.20} each operator in the cluster has an incoming 
and an outcoming arrow. 
Certainly, since $n_c=n_{c'}$, the 
product $W^{(\ell')} \cdot W^{(\ell)}$ will also appear in the summation 
\rf{2.12}. We  can obtain $W^{(\ell')}$ by exchanging all the 
non-commuting generators of $W^{(\ell)}$ by the ones in $W^{(\ell')}$. 
The addition of both contributions vanishes, i. e.,  
$[W^{(\ell)},W^{(\ell')}] + 
[W^{(\ell')},W^{(\ell)}] =0$.

Consider now the cases where we have in the product $W^{(\ell)}\cdot W^{(\ell')}$ a single connected cluster with $n_c \neq n_{c'}$ generators in $W^{(\ell)}$  
and $W^{(\ell')}$, respectively. We call then connected odd clusters. The 
clusters ($c$) and $(d$) in Fig.~\ref{Q5} are examples of them. 
Since $n_{c'} = n_c \pm 1$, necessarily $n_c$ or $n_{c'}$ is even. If $n_c$ 
is odd ($n_{c'}$ is odd) the generators in $W^{(\ell)}$ (in $W^{(\ell')}$) 
will have a pair of non-commuting generators in $W^{(\ell')}$, respecting 
the arrow condition \rf{2.19}
(in $W^{(\ell)}$), implying $[W^{(\ell)},W^{(\ell')}]=0$. 

In general the cluster configuration in the product 
$W^{(\ell)}\cdot W^{(\ell')}$ will have an arbitrary number of even and odd 
connected clusters. If we now consider the related words $\tilde{W}^{(\ell)}$ 
and $\tilde{W}^{(\ell')}$ where we interchange only the generators belonging 
to the even connected cluster, we have 
\be \label{2.21}
[W^{(\ell)},W^{(\ell')}] +[\tilde{W}^{(\ell)},\tilde{W}^{(\ell')}]=0,
\ee
since the generators in $W^{(\ell)}$ and $W^{(\ell')}$ forming the odd 
connected clusters commute.

To summarize if  the generators $\{h_i\}$, defined in \rf{2.19}, do not have 
a single generator $h_k$ that does not commute with three (or more) 
 commuting generators, the charges defined in \rf{2.11} are in involution 
\rf{2.12}. As $M \to \infty$ the number of charges $\ell$ also 
diverge, and 
we have the exact integrability of the model. In the case where 
$h_k$ do not commute with two operators, $h_m$ and $h_n$, they should 
satisfy
\be \label{2.22} 
h_k\;h_m = \rho h_m\;h_k, \quad h_k\;h_n = \rho^{-1} h_n\;h_k.
\ee

It is interesting to stress that the involution, or exact integrability holds 
for arbitrary $\rho \in \mathbb{C}$. For  $\rho$ arbitrary, $h_i^M, M=1,2,\ldots$ are 
distinct and linearly independent words and the algebra actually have an infinite number of 
generators, even for 
a finite number of $\{h_i\}$. In the special cases where $\rho = e^{i2\pi/N}$ 
($N=2,3,4,\ldots$), the number of generators is finite for a finite 
number of $\{h_i\}$, and we have in \rf{2.19} and \rf{2.22} a $Z(N)$-exchange 
algebra.

\begin{figure} [htb]
\centering
\includegraphics[width=0.45\textwidth]{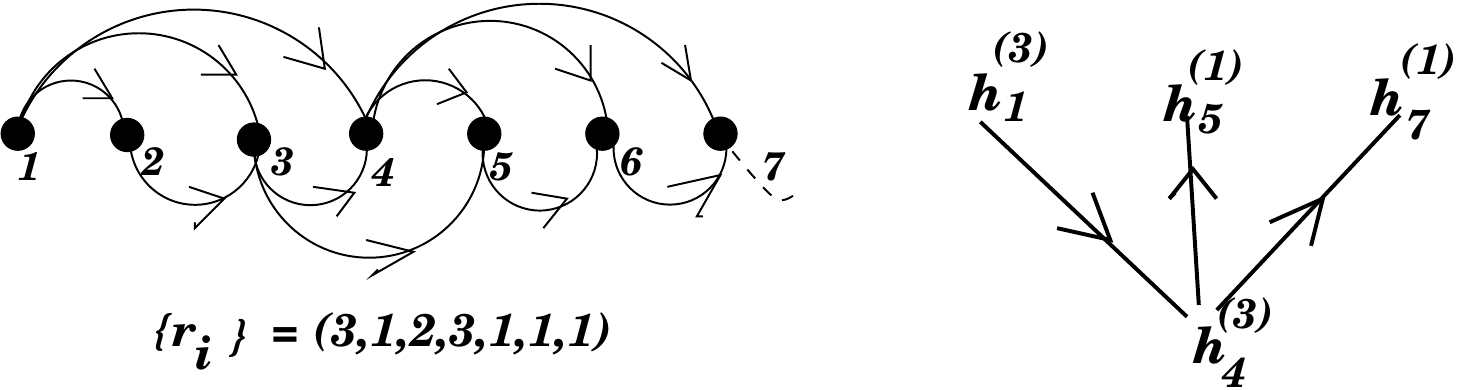}
\caption{
The $M=7$ generators in the Hamiltonian \rf{2.1}.
The circles represent the 
generators $h_1^{(3)},h_2^{(1)},h_3^{(2)},$ $h_4^{(3)},h_5^{(1)},h_6^{(1)},
h_7^{(1)}$.
The links in the diagram connect the non-commuting generators. The generator
 $h_4^{(3)}$ does not commute with the commuting generators 
$h_1^{(3)},h_5^{(1)}$ and $h_7^{(1)}$ (see the diagram in the right).} 
\label{Q6}
\end{figure}

\begin{figure} [htb]
\centering
\includegraphics[width=0.45\textwidth]{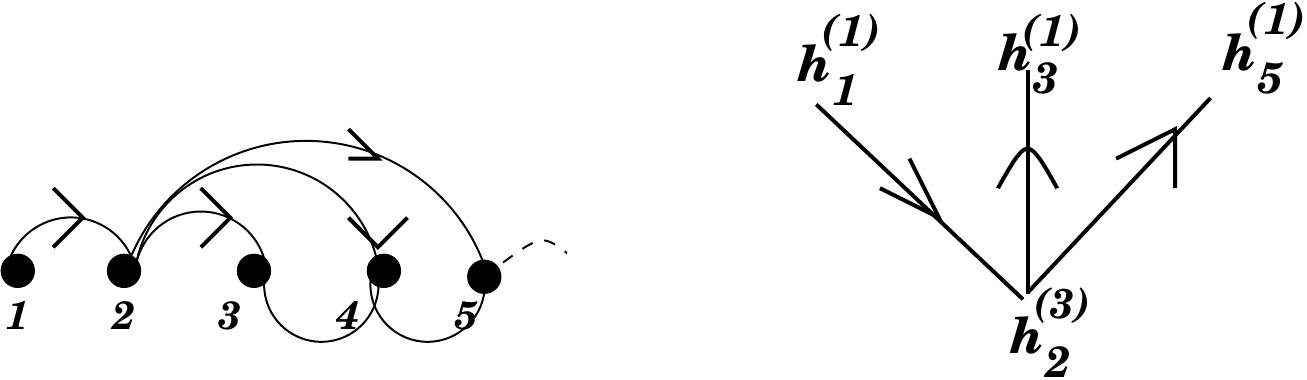} 
\caption{
Representations of the generators $h_1^{(1)},h_2^{(3)},h_3^{(1)},h_4^{(1)},
h_5^{(1)}$. The links in the diagram connect the non-commuting generators. 
The generator $h_2^{(3)}$ does not commute with the generators 
$h_1^{(1)},h_3^{(1)}$ and $h_5^{(1)}$ (see the diagram in the right).
} \label{Q7}
\end{figure}

We now return to the one-dimensional general models where the 
generators $h_i^{(r_i)}$ 
satisfy \rf{2.2a}-\rf{2.2b}. Let us consider some examples. The Hamiltonian \rf{2.1} 
with $M=7$ generators 
$h_1^{(3)},h_2^{(1)},h_3^{(2)},h_4^{(3)},h_5^{(1)},h_6^{(1)},h_7^{(1)}$ 
is shown in Fig.~\ref{Q6}.
In the figure \ref{Q7}  we also show the link diagram where the non-commuting generators 
are linked. In this case the generator $h_4^{(3)}$ does not commute 
with the commuting operators $h_1^{(3)},h_5^{(1)}$ and $h_7^{(1)}$, and then 
the algebra  do not satisfy the constraint ensuring the involution \rf{2.12} 
(see Fig.~\ref{Q2}). 
Another 
example is the case where we have the generator $\{h_i^{(r_i)}\}$ where 
$r_1,r_2,r_3,r_4,r_5=1,3,1,1,1$ as in Fig.~\ref{Q7}. We see that the 
generators $h_2^{(3)}$ does not commute with the commuting operators 
$h_1^{(1)},h_3^{(1)}$ and $h_5^{(1)}$, and this set of ranges is not allowed.

It is important to observe that in our $Z(N)$ exchange algebra \rf{2.4} the 
generator $h_i^{(r_i)}$, does not commute with all the generators 
$h_j{(r_j)}$, on its right with $i<j\leq i+r_i$, and all arrows linking 
these sites with the site $i$ {\it have the same orientation}.

Consider the pair of generators $h_i^{(r_i)},h_{i+1}^{(r_{i+1})}$ 
($r_i>0$), that do not commute. The 
generator  $h_j^{(r_j)}$ also do not commute 
with $h_i^{(r_i)}$ if $j\leq i+r_i$, but commutes with $h_{i+1}^{(r_{i+1})}$  
 if $j\geq(i+1) +r_{i+1}+1$. 
This means that, if $0\leq r_{i+1} \leq r_i-2$, we have a violation 
of condition \rf{2.19}, because the link orientation of non-commuting of 
$h_i^{(r_i)}$ with 
$h_{i+1}^{(r_{i+1})}$ and $h_j^{(r_j)}$ are the same. 
In the case where $r_{i+1} \geq r_i -1$ the condition \rf{2.19} is not 
violated because $h_{i+1}^{(r_{i+1})}$ although not commuting with 
$h_i^{(r_i)}$ also does not commute with all the generators $h_j^{(r_j)}$ 
with $i+1< j \leq i+r_i$.

In summary, for the general parafermionic models (also for $\rho$ general)
 the interacting ranges should be restricted in a solid-on-solid (RSOS) 
 path where 
\be \label{2.23}
r_{i+1} \geq r_i -1, \quad (i=1,\ldots,M).
\ee

In the fermionic case, since $\rho={\rho}^{-1}=1$, the condition \rf{2.19} 
is less restrictive, as compared with the $Z(N)$ parafermionic cases. In this 
case there are some additional cases that violates \rf{2.23}. 
In special consider the cases where the first generator $h_1^{(r_1)}$ does 
not commute with $h_2^{(r_2)},h_{i+r_1}^{(r_{i+r_1})}$ and 
$h_{2+r_2+1}^{(r_{2+r_2+1})}$. These three last operators commute 
among them, if 
\be \label{newq}
r_{2+r_2+1} <(1+r_1)-(2+r_2+1)-1 = r_1-r_2 -3,
\ee
which is not allowed due to the condition \rf{2.20} (see figure~\ref{Q2}).
On the contrary if $r_{2+r_2+1} \geq r_1-r_2 -3$ we do not have a violation 
of the constraint \rf{2.19}, since $h_{2+r_2+1}^{(2+r_2+1)}$ and 
$h_{1+r_1}^{(r_{1+r_1})}$ do not commute. In particular we may have 
free-fermionic (not free-parafermionic) models with generators 
$h_1^{(r)},h_2^{(r)},\dots,h_k^{(r)},h_{k+1}^{(r')},h_{k+2}^{(r')},\dots 
h_M^{(r')}$, if $(r-2r') \geq 3$ and $k=0,1,2,\ldots$. 


 \subsubsection{ The recursion relation for the charges }

\begin{figure} [htb]
\centering
\includegraphics[width=0.45\textwidth]{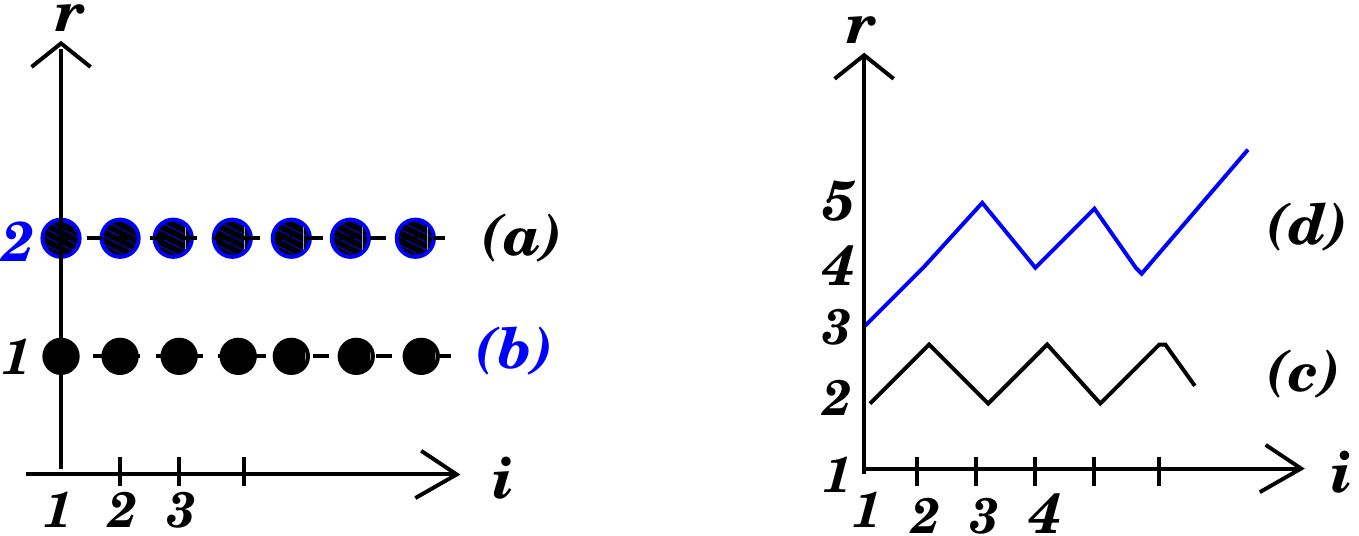} 
\caption{
Examples of allowed RSOS paths, for free-particle models. In (a) and (b) we have  homogeneous interacting range models, where  the ranges have the values 
1 and 3, respectively. In (d) and (c) we have examples of models with 
non-homogeneous range of multispin interactions.} \label{Q8}
\end{figure}

In Fig.~\ref{Q8} we show examples of the RSOS path \rf{2.23}. The path 
($a$) and ($b$) are the ones of the homogeneous models in \cite{AP1,AP2} and 
the path ($c)$ and ($d$) are examples of inhomogeneous ones.

It is convenient to observe that due to the restriction \rf{2.23} we can also 
obtain the number of commuting generators $\{\ell_i\}$ at left of the 
generators $\{h_i^{(r_i)}\}$. For $i=2,3,\ldots$ :
\be \label{2.24}
\ell_i = \mbox{Max}_{0<j<i}\{i-j\}, \quad \mbox{where} \quad r_j\geq (i-j)>0.
\ee
 Also
 $l_1=0$ since we do not have any generator at left of the site $i=1$. 

We can redefine the generators $h_i^{(\ell_i,r_i)} \equiv h_i^{(r_i)}$, 
and the relations \rf{2.2a} are now  given by: 
\be 
h_i^{(\ell_i,r_i)} h_j^{(\ell_j,r_j)} = \begin{cases} 
\frac{1}{\omega} h_j^{(\ell_j,r_j)} h_i^{(\ell_i,r_i)},&0<(i-j)\leq \ell_i, \\
\omega h_j^{(\ell_j,r_j)} h_i^{(\ell_i,r_i)}, &0<(j-i)\leq r_i, \\
h_j^{(\ell_j,r_j)} h_i^{(\ell_i,r_i)}, & \mbox{otherwise}. 
\nonumber
\end{cases}
\ee

In terms of $\{h_i^{(\ell_i,r_i)}\}$ it is not difficult to derive the 
recursion relation for the conserved charges $Q_M^{(\ell)}$ of the quantum
chain given in \rf{2.1}
\bea \label{2.26}
Q_M^{(1)} &=& \sum_{i=1}^M h_i^{(\ell_i,r_i)}, \nonumber \\
Q_M^{(\ell)} &=& Q_{M-1}^{(\ell)} + 
h_M^{(\ell_M,r_M)} Q^{(\ell-1)}_{M-(\ell_M+1)},
\eea
with the initial conditions $Q_M^{(0)}=1$ for $M\neq 0$ and $Q_0^{(\ell)}=0$ 
for $\ell >0$.
The last relation is a generalization of the recursion relation given in 
\cite{AP1,AP2} for the homogeneous models.

\subsubsection{Free-particle spectra}

The question that remains concerns the verification that the integrable 
models \rf{2.1}, whose generators satisfy \rf{2.2a},\rf{2.2b} and \rf{2.23} have a 
free-particle spectra. 
A sufficient condition, as verified in the known homogeneous case, where 
$r_1=r_2=\cdots $, is the existence of a closure relation 
(inverse relation) for the charge generating function
\be \label{2.27}
G_M(u) = \sum_{\ell=0}^{\bar{M}} (-u)^{\ell} Q_M^{(\ell)},
\ee
where $\bar{M}$ is the number of independent $Q_M^{(\ell)}$ charges. 

In the homogeneous case this closure relation is given by the special 
$Z(N)$ product:
\bea \label{2.28}
&&T_M(u) = \prod_{j=0}^{N-1} G_M(\omega^j u) 
= \sum_{\ell=1}^{\bar{M}} u^{\ell} \nonumber \\
&&\left( 
\sum_{\ell_1,\ldots,\ell_N}^{(*,\ell)} 
Q_M^{(\ell_1)} \cdots Q_M^{(\ell_M)} \omega^{0\ell_1 + 
1\ell_2 + \cdots +(N-1)\ell_N}\right),
\eea
where the restriction $(*,\ell)$ in the summation means 
$\sum_{i=1}^N \ell_i = \ell$.
Since the charges commute among themselves, for a given set 
($\ell_1,\ell_2,\ldots,\ell_N$), that $\sum_{i=1}^N \ell_i=\ell$, the 
contribution in \rf{2.28} is proportional to 
\be \label{2.29a}
\sum_{\{P\}} \omega^{\ell_{P_2}+2\ell_{P_3}+ \cdots +(N-1)\ell_{P_N}},
\ee
where $P=(P_1,P_2,\ldots,P_N)$ are the $N!$ permutations of the 
integers $1,2,\ldots,N$. However \rf{2.29a} gives zero if $\sum_{i=1}^N\ell_i=\ell$ is not an integer multiple of $N$. This imply that only powers of $u^N$ appears 
in \rf{2.28}. Remarkable, for the homogeneous systems, the coefficients of 
$u^{\ell}$ are just c-numbers. i. e.,
\be  \label{2.29b}
T_M(u) = \sum_{\ell=0}C_M^{(\ell)} u^{\ell N} = P_M(u^N),
\ee
and $G_M(u)$ is a polynomial $P_M(z)$ of the variable $z=u^N$.

The relation \rf{2.28} ensure that all the conserved charges $Q_M^{(\ell)}$ 
have a free-particle spectra. In order to see this we notice that 
$[Q_M^{(\ell)},Q_M^{(\ell')}] =0$ imply also $[G_M(u),G_{M}(u')]=0$, and all the 
charges, as well $G_M(u)$ share the same eigenfunctions $\ket{\Psi}$. 
Denoting the eigenvalue of $G_M(u)$ by $\Lambda_M(u)$, we now have from
 \rf{2.28} 
\bea \label{2.29}
\Lambda_M(u)\Lambda_M(\omega u) \cdots \Lambda_M(\omega^{N-1}u) &=& P_M(u^N) \nonumber \\ 
&=&\prod_{i=1}^{\bar{M}} \left( 1- \frac{u^N}{z_i}\right),
\eea
since $P_M(0)=1$ and $z_i$ ($i=1,2,\ldots,\bar{M}$) are the roots of the 
polynomial $P_M(z_i)=0$. Rewriting 
\be \label{2.30}
1-\frac{u^N}{z_i} = \prod_{j=0}^{N-1}
\left(1-u\frac{\omega^j}{z_i^{1/N}}\right),
\ee
we obtain the possible ${\bar{M}}^N$ solutions for $\Lambda_M(u)$:
\be \label{2.31}
\Lambda_M(u)^{\{s_i\}} = \prod_{i=1}^{\bar{M}} 
\left( 1 - u\frac{\omega^{s_i}}{z_i^{1/N}}\right) 
= \prod_{i=1}^{\bar{M}} 
\left( 1- u\omega^{s_i} \varepsilon_i\right) , 
\ee
where
$\varepsilon_i = 1/z_i^{1/N}$, and for each root $z_i$ we choose one of the 
possibilities $s_i = 0,1,\ldots,N-1$.

Expanding the generator $G_M(u)$ and $\Lambda_M(u)$ in powers of $u^N$ we 
find the possible eigenvalues of all the charges $Q_M^{(\ell)}$. In particular 
the eigenvalues of the Hamiltonian $-Q_M^{(1)}$ have the free-particle spectra
\be  \label{2.32}
E_{s_1,\ldots,s_{\bar{M}}} = 
- \sum_{i=1}^{\bar{M}} \omega^{s_i} \varepsilon_i,
\ee
For each root $z_i$ (pseudo-energy $\varepsilon_i=1/z_i^{1/N}$), given in \rf{2.31}, we have to
 chose {\it{one and only one}} phase $\omega^{s_i}$, to form the eigenenergy 
$E_{s_1,\ldots,s_{\bar{M}}}$. This is the $Z(N)$ circle 
exclusion constraint (see 
\cite{baxter1,baxter2, fendley1,baxter3,perk1,perk2,AB1,AB2}), i.e.,  
the parafermionic version of the Pauli exclusion principle. 

In appendix A we show that for the general models \rf{2.1}, given in terms 
of generators of the $Z(2)$ exchange algebra \rf{2.2a}-\rf{2.2b}, the inversion 
relation 
\be \label{2.33}
G_M(u)G_M(-u) = \sum_{\ell}{\cal{C}}_M^{(\ell)}\;u^{2\ell} = P_M(u^2),
\ee
is satisfied  provide the following two conditions are verified. 

{\bf{(a)}} 
{\it There is no single generator that does not commute with three 
other commuting operators.} 

{\bf{(b)}}
{\it By linking the generators that do not commute we do not form closed 
loops (see appendix A).}

The condition (b) is satisfied for the quantum chains with open boundary 
conditions, but exclude the inverse relation \rf{2.33} for periodic chains. 
The proof of the inversion relation \rf{2.29b} for $N>2$, on a 
direct way, as we did in appendix A for $N=2$, is cumbersome, but it was 
already obtained in \cite{ref-inv-zn} using graph theory.

The polynomial $P_M(u^N)$, for general $N$, has the coefficients ${\cal{C}}_M^{(\ell)}$ in \rf{2.29b}, given by (see appendix A)
\be \label{2.34}
{\cal{C}}_M^{(\ell)}= (-)^{\ell}\sum^{*}_{\{i_1,i_2,\ldots,i_{\ell}\}} 
\lambda_{i_1}^N \lambda_{i_2}^N \cdots \lambda_{i_{\ell}}^N,
\ee
where now the symbol (*) denotes the sum over all independent possibilities 
of the products of the $\ell$-couplings associated to the $\ell$-commuting 
generators 
$h_{i_1}^{(r_1)}h_{i_2}^{(r_2)} \cdots h_{i_{\ell}}^{(r_{\ell})}$ in the 
charge $Q_M^{(\ell)}$.
 
The recursion relation \rf{2.26} of the charges $Q_M^{(\ell)}$ give us a 
recursion for the coefficients ${\cal{C}}_M^{(\ell)}$:

\be \label{2.34}
{\cal{C}}_M^{(\ell)} = 
{\cal{C}}_{M-1}^{(\ell)} - \lambda_M^{N}\;
{\cal{C}}_{M-(\ell_M+1)}^{(\ell-1)} \mbox{ for  } M>1,
\ee 
where ${\cal{C}}_0^{(0)} =1$ and ${\cal{C}}_{j'}^{(\ell')} =0$ if $\ell'>0$ 
or $j'<0$.

The above relation imply the recursion for the polynomial
\be \label{2.36}
P_M(z) = P_{M-1}(z) - z\;\lambda_M^N P_{M-(l_M+1)}(z), M\geq 1,
\ee
with the initial condition $P_M(z)=1$ for $M\leq0$.

\section{Examples of models with inhomogeneous interacting ranges}

 In the last section we verified that arbitrary one-dimensional quantum chains 
\be \label{3.1} 
H_M^{(N,\{r_i\})}(\lambda_1,\ldots,\lambda_M) = - \sum_{i=1}^M h_i^{(r_i)},
\ee
with the site-dependent interaction ranges $\{r_i\}$, are exactly integrable 
and have a free-particle eigenspectra if the generators $\{h_i^{(r_i)}\}$ 
obey the $Z(N)$ algebra \rf{2.2a}-\rf{2.2b}, with the restriction 
\be \label{3.2}
r_{i+1} \geq r_i -1 \quad (i=1,\ldots,M).
\ee

Arbitrary representations of the algebra \rf{2.2a}-\rf{2.2b} 
satisfying \rf{3.2} 
will give us integrable quantum chains \rf{3.1} with a free-particle spectra 
\rf{2.32}.

A particular representation of dimension $N^M$, that we can always construct, 
is the one we call "word representation". 
In this representation we call the generators of the algebra as "letters" 
(alphabet) and the product of them as "words". The vector space is spanned
 by the vectors $|s_1,s_2,\ldots,s_M>$, ($s_i = 0,1,\ldots,N-1$), with a 
one-to-one correspondence with the words
\be \label{3.3}
|s_1,s_2,\ldots,s_M> \leftrightarrow [h_1^{(r_1)}]^{s_1} \dot 
 [h_2^{(r_2)}]^{s_2} 
\cdots   [h_M^{(r_M)}]^{s_M} . \nonumber
\ee
In this representation the generators are given by the generalization of 
\rf{2.8}
\be \label{3.4}
h_i^{(r_i)} = \lambda_i \left( \prod_{j=i-\ell_i}^{i-1} Z_j\right) X_i,
\ee
where $\{\ell_i\}$ are obtained from $\{r_i\}$ as in \rf{2.24}. 
The $M$-sites Hamiltonian is given by
\be \label{3.5} 
H_M^{(N;\{r_i\})} (\lambda_1,\ldots,\lambda_M) = 
-\sum_{i=1}^M\lambda_i \left( \prod_{i=i-\ell_i}^{i-1} Z_j \right) X_i. 
\ee
The homogeneous cases \rf{2.3} is recovered by taking 
$r_1=r_2=\ldots,r_M=p$ ($p=1,2,\ldots$), where we have the $p$-multispin 
free fermionic $N=2$ and free parafermionic $(N>2$) models.

\subsection{A simple example}

\begin{figure} [htb]
\centering
\includegraphics[width=0.30\textwidth]{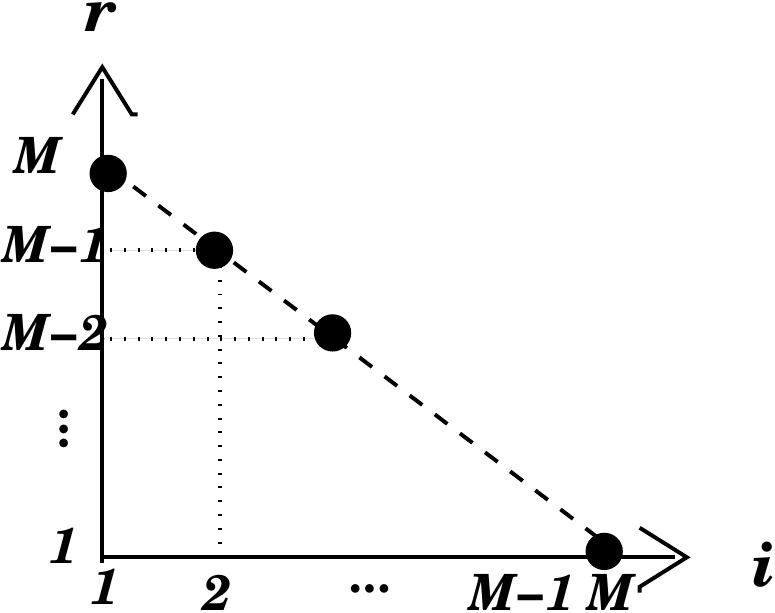} 
\caption{
Multispin interacting ranges for a simple model where $r_i=M+1-i$ 
($i=1,\ldots,M$). The model only have one conserved charge 
$Q_M^{(1)}=-H$.} \label{exo1}
\end{figure}

A simple and exotic example is the $M$-sites chains given in terms of $Z(N)$ 
generators where the interaction ranges  
$ r_i = M-i+1$, for $i=1,\ldots,M$ as in Fig.\ref{exo1}.

In this case, no matter how large is $M$ we only have the $\ell$-charge with 
$\ell=1$, i. e.,
\bea \label{3.6} 
Q_M^{(1)} &=& -H_M^{(\{r_i\})} = \nonumber \\
&&\lambda_1h_1^{(M)} + \lambda_2h_2^{(M-1)} + \cdots 
+\lambda_Mh_M^{(1)},
\eea
with $[h_i^{(1)}]^N =1$.
Since now $G_M(u)=1+uH_M^{(\{r_i\})}$, is simple to see that
\be \label{3.7} 
T_M(u) = \prod_{j=0}^{N-1} G_M(\omega^ju)=P_M(u) = 
1-u^N \left(\sum_{i=1}^M \lambda_i^N\right), \nonumber
\ee
that gives $P_M(z)=1-z\sum_{i=1}^M \lambda_i^N$.
There is a single root of $P_M(\bar{z})=0$, $\bar{z} = \sum_i^N\lambda_i^N$ if 
$\sum_i^N \lambda_i^N \neq 0$, and there is only $N$ non-zero energies in the 
Hamiltonian 
\be \label{3.8}
\varepsilon_s = e^{i\frac{2\pi}{N}s} /\left(\sum_{i=1}^M \lambda_i^N\right), \quad s=0,1\ldots,N-1.
\ee
The $N^M$-dimensional Hilbert space has the above eigenenergies  with 
degeneracy $N^M/N= N^{M-1}$.

In the case where $\sum_i^M \lambda_i^N=0$, the polynomial is $P_M=1$ and 
we have no zeros. This means that all the eigenenergies are zero. If $H_M^{(\{r_i\})}$ 
would be diagonalizable it should be a zero matrix. Since this is not the 
case $H_M^{(\{r_i\})}$ is not fully diagonalizable and have a Jordan-cell structure. 

Actually in this exotic example all the generators do not commute, obeying 
a $Z(N)$ algebra. In the case $N=2$ they satisfy 
\be \label{3.9}
\{h_i^{(r_i)},h_j^{(r_j)}\} = 2\delta_{i,j}, \quad (h_i^{(r_i)})^2 =1,
\ee
and we can identify them as Majorana fermions. The 
Hamiltonian \rf{3.6} is just the sum of Majorana fermions. 

In the case $N>2$ we can identify the generators as a generalized 
$Z(N)$ Majorana parafermions \cite{zn-parafermions}, where 
\be \label{3.10}
h_i^{(r_i)} h_j^{(r_j)} = \begin{cases} 
e^{i\frac{2\pi}{N}}  h_j^{(r_j)} h_i^{(r_i)},& j>i \\
e^{-i\frac{2\pi}{N}}  h_j^{(r_j)} h_i^{(r_i)},& j<i , 
\end{cases}
\ee
and $(h_i^{(r_i)})^N =1$.

\begin{figure} [htb]
\centering
\includegraphics[width=0.30\textwidth]{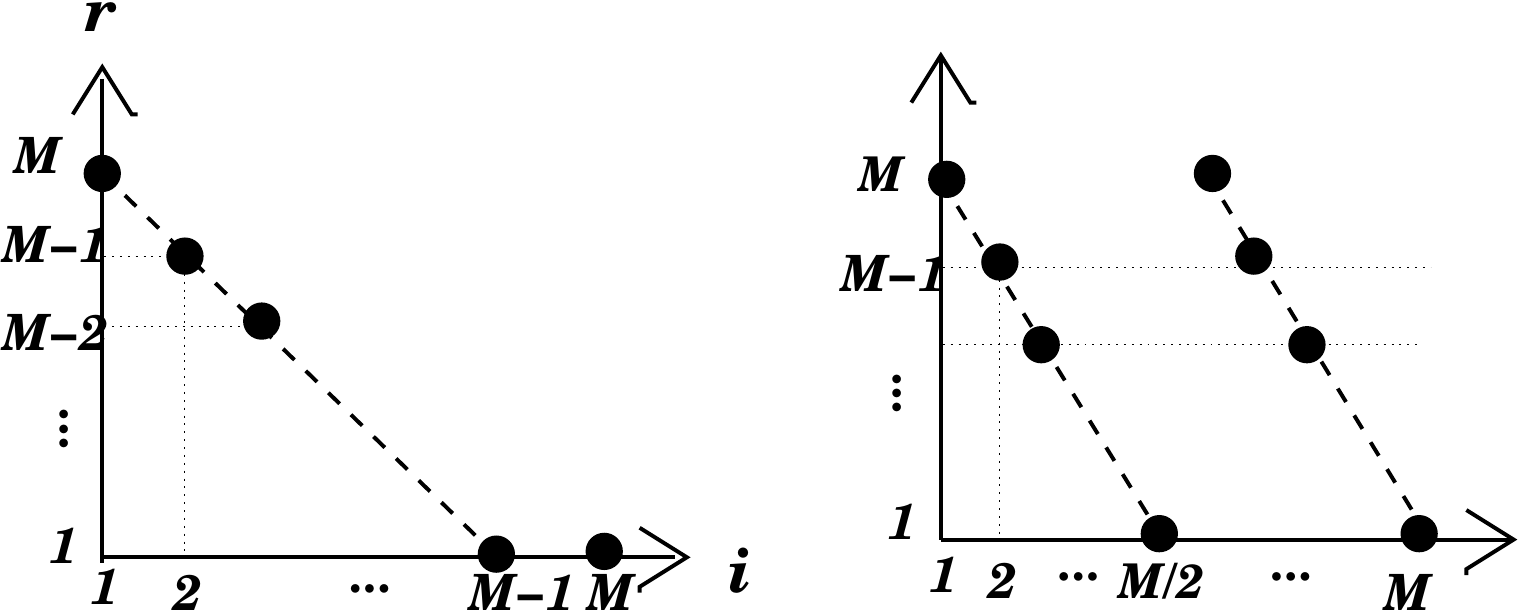} 
\caption{
Examples of two simple models where the only conserved charges are 
	$Q_M^{(1)}=-H$ and $Q_M^{(2)}$.} \label{exo2}
\end{figure}

Many more interesting examples can be produced where the range of 
interactions obey the restriction \rf{3.2}. In Fig.~\ref{exo2} we show the range of the multispin interactions for two other simple examples.
In this case we only have the conserved charges $Q_M^{(1)}= -H$ and $Q_M^{(2)}$. 
The related polynomial in these cases are of second order. 

\subsection{Models where all the even and odd sites have a constant range of 
interactions}

The Hamiltonians are given by 
\be \label{3.11}
H= -\sum_{i=1}^{M/2} (\lambda_o h_{2i-1}^{(r_o)} + \lambda_e h_{2i}^{(r_e)}),
\ee
where the generators $\{h_{2i-1}^{(r_o)},h_{2i}^{(r_e)}\}$ satisfy the 
$Z(N)$ algebra \rf{2.2a}, and $M$ is even.  For convenience the 
generators instead of satisfying \rf{2.2b} now satisfy $[h_i^{(r_i)}]^N=1$.

We can consider initially the case where $r_o=3$ and $r_e=2$, i. e., the 
interacting ranges are ($3,2,3,2,\ldots$).
In this case by adding pairs of generators we define new ones
\be \label{3.12}
\tilde{h}_i^{(1)} = \lambda_o h_{2i-1}^{(3)} + \lambda_e h_{2i}^{(2)},
\ee
and the Hamiltonian \rf{3.11} is now given by
\be \label{3.13}
 H =  -  \sum_{i=1}^{M/2} \tilde{h}_i^{(1)}.
\ee
It is simple to verify from \rf{2.2a}-\rf{2.2b} that the new generators satisfy
\be \label{3.14} 
\tilde{h}_i^{(1)} \tilde{h}_j^{(1)} = \begin{cases} 
e^{i\frac{2\pi}{N}}  \tilde{h}_j^{(1)} \tilde{h}_i^{(1)},& j=i+1 \\
  \tilde{h}_j^{(1)} \tilde{h}_i^{(1)},& j>(i+1),
\end{cases}
\ee
with  
\be \label{3.14a}
[\tilde{h}_{i}^{(1)}]^N=[\tilde{\lambda}_i^{(1)}]^N =
[\lambda_o h_{2i-1}^{(3)} + \lambda_e h_{2i}^{(2)}]^N = 
\lambda_{o}^N + \lambda_e^N. \nonumber
\ee
In the derivation of this last formula we use the algebraic relations 
\rf{3.13} and the fact that $1 + \omega + \cdots +\omega^{N-1} =0$.

This last algebra is the one of the fermionic and parafermionic 
uniform range models where $r_1=r_2=\ldots=p=1$. The model is then 
in a multicritical point  
\cite{AP1,AP2} for arbitrary values of $\lambda_o$ and $\lambda_e$, with 
dynamical critical exponent $z= 2/N$.

This result is promptly generalized for the models where the ranges in 
\rf{3.11} are 
\be \label{3.53a}
r_{e}=\ell, \quad r_o = \ell+1, \quad \mbox{for  } 0<\ell,   \quad \mbox{even }.
\ee
In this case we extend \rf{3.12} by defining 
\be \label{3.16}
\tilde{h}^{(p)} = \lambda_oh_{2i-1}^{(\ell+1)} +\lambda_e h_{2i}^{(\ell)}, \quad p=\frac{\ell}{2},
\ee
and as before we obtain the Hamiltonian \rf{3.13} with generators 
$\{\tilde{h}_i^{(p)}\}$. We can verify that 
\be \label{3.17}
\tilde{h}_i^{(p)} \tilde{h}_j^{(p)} = \begin{cases} 
e^{i\frac{2\pi}{N}}  \tilde{h}_j^{(p)} \tilde{h}_i^{(p)},& 0<(j-i)\leq p \\
  \tilde{h}_j^{(p)} \tilde{h}_i^{(p)},& (j-i)>p,
\end{cases}
\ee
with 
$(\tilde{h}_i^{(p)})^N= \lambda_o^N + \lambda_e^N$, and $p = \ell/2$. The model is now equivalent 
to an uniform multispin interacting range $p = \ell/2$ model. The model 
is in a multicritical point for arbitrary values of $\lambda_o$ and 
$\lambda_e$, having a dynamical critical exponent $z = (p+1)/N= (\ell+2)/(2N)$ \cite{AP1,AP2}.

We consider now the cases where 
\be \label{3.18}
r_e=\ell, \quad r_o=\ell+1 \quad \mbox{for }  \ell \mbox{ odd}.
\ee
In these cases the transformation \rf{3.16} give operators that do not 
satisfy the algebra \rf{2.2a}-\rf{2.2b}.
\begin{figure} [htb]
\centering
\includegraphics[width=0.30\textwidth]{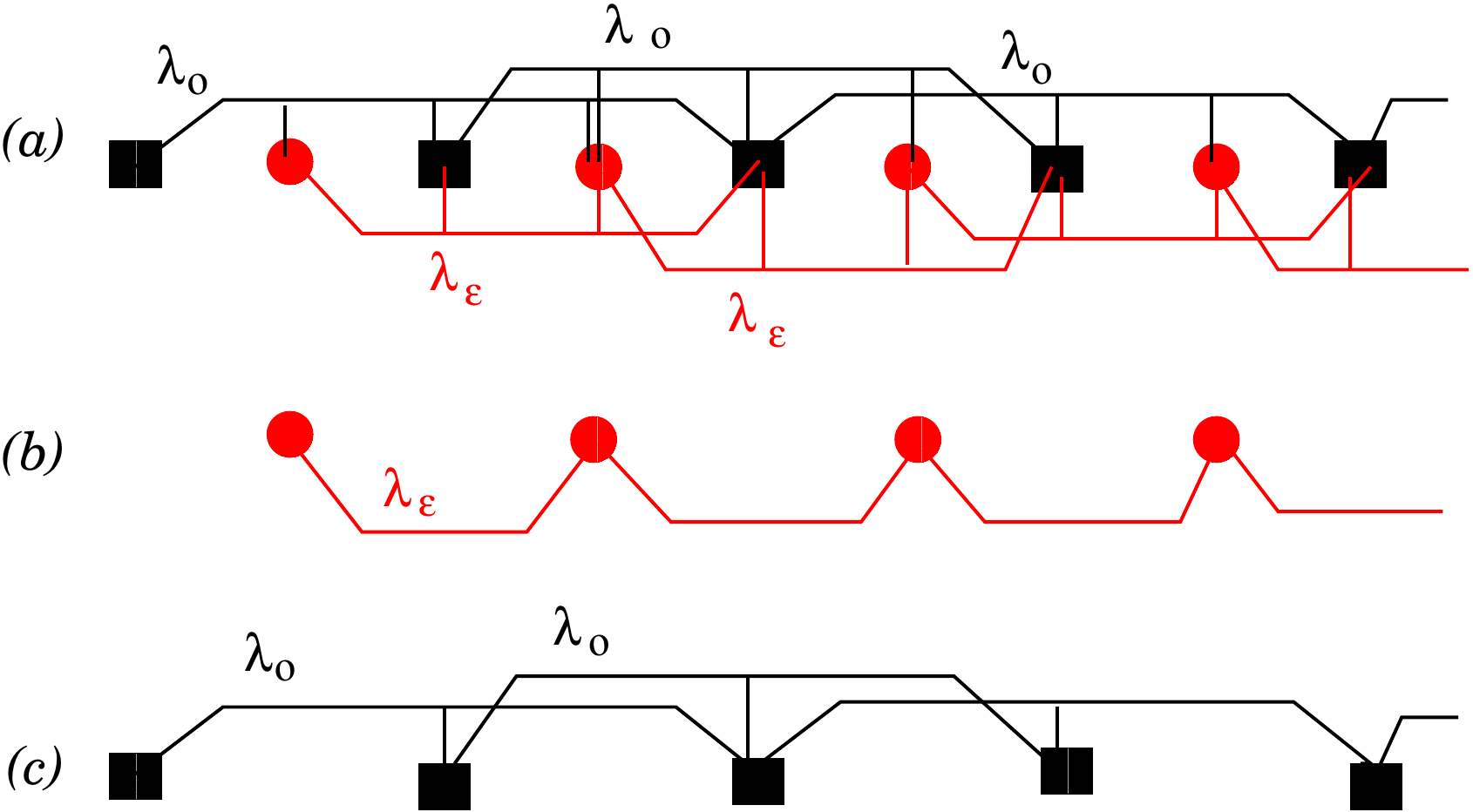} 
\caption{
Representation of the quantum chains with multispin coupling constants 
$\lambda_o$ and $\lambda_e$ in the odd and even sites, respectively. In the 
figure (a) it is shown the case of the model with 
  multispin interacting ranges $r_o=4$ and 
$r_e=3$, for the odd and even sites. In (b) we represent the limiting 
case where $\lambda_o=0$, and we have an effective homogeneous interacting 
range model with two-spin interactions. In (c) we have the limiting case where 
$\lambda_e=0$. In this case we have an effective model where  all  multispin 
interactions  are uniform, having range $r_o=3$.} \label{aux1}
\end{figure}

Let us see some limiting cases. If $\lambda_o=0$ and $\ell>1$ in \rf{3.11} 
 (see Fig.~\ref{aux1}a,b), we end up with a model $H=-\lambda_e\sum_{i=1}^{M/2}\tilde{h}_i^{(\frac{\ell-1}{2})}$, where the generators  $\tilde{h_i}^{(\frac{\ell-1}{2})}$ satisfy the 
algebra \rf{2.2a}-\rf{2.2b} with an uniform range of interaction $p=(\ell-1)/2$. 
For $\ell >1$ the 
Hamiltonian in this limit is in a multicritical point for arbitrary values of
$\lambda_e$, having a dynamical critical exponent $z=(\ell +1)/(2N)$ \cite{AP1,AP2}.

For $\ell=1$ the resulting model is a set of non-interacting spins, since 
$p=0$, being gapped.
 
The other limit where $\lambda_e=0$ (see Fig.~\ref{aux1}a,c) give us the Hamiltonian 
$H=-\lambda_o\sum_{i=1}^{M/2} \tilde{h}_i^{(\frac{\ell+1}{2})}$, with  generators satisfying 
the algebra \rf{2.2a}-\rf{2.2b} with an uniform interacting range $p =(\ell+1)/2$. 
For $\ell>2$ and arbitrary values of 
$\lambda_o$, it is in a multicritical point with dynamical critical exponent 
$z=(\ell+3)/(2N)$. 

Apart from these limiting cases our analysis will be done numerically. The 
mass gaps of the Hamiltonians  with $M$ sites are obtained from the largest 
root of the polynomial $P_M(z)$, generated by the recursion relation given 
in \rf{2.36}. 

\begin{figure} [htb]
\centering
\includegraphics[width=0.30\textwidth]{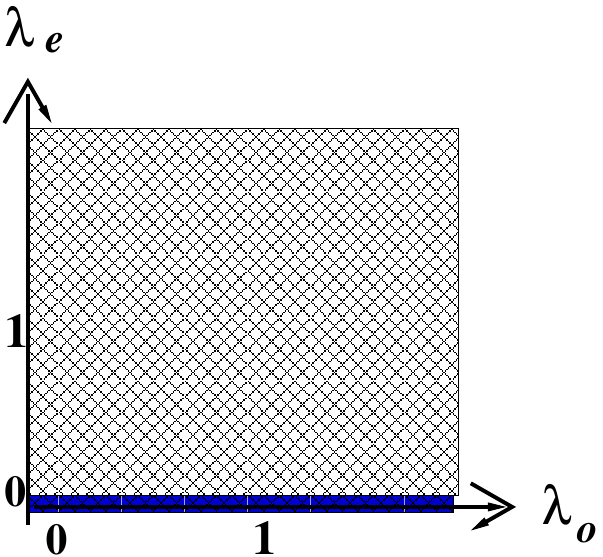} 
\caption{
Schematic representation of the phase diagram of the models described by
the Hamiltonian \rf{3.11}. The models are critical in general, for all values 
of $\lambda_o$ and $\lambda_e$. They belong to  the same universality class in 
the whole plane ($\lambda_o,\lambda_e$), except in the line where 
$\lambda_e=0$. The case where $r_o=2$ and $r_e=1$, is exceptional. The model 
is gapped in the whole plane, except at the line $\lambda_e=0$.}\label{f1-xfig}
\end{figure}

Our studies indicates the schematic 
phase diagram  shown  in Fig.~\ref{f1-xfig}
For $\ell=3,5,7$ and $9$ the model is always critical in the parameter 
space ($\lambda_o,\lambda_e$). 
In general the model is in a critical universality class with dynamical 
critical exponent $z=(\ell+1)/(2N)$, except for the line where 
$\lambda_e =0$ where $z = (\ell+3)/(2N)$. 
In the case $\ell=1$, the model is massive in the whole phase diagram except 
 when $\lambda_e=0$, where the dynamical critical exponent is $z=2/N$. 

\begin{figure} [htb]
\centering
\includegraphics[width=0.45\textwidth]{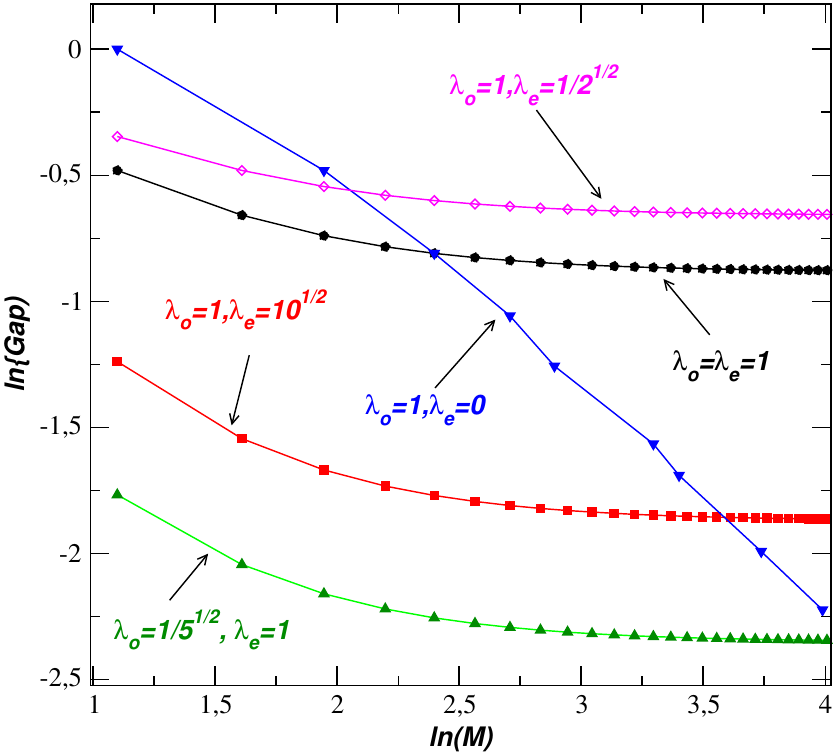} 
\caption{
The finite-size gaps for the Hamiltonian \rf{3.11} with range of multispin interactions $r_o=2$ and $r_e=1$. The lattice sizes are  $M \leq 60$.
The gaps are calculated for several values of $\lambda_o$ and $\lambda_e$ in 
the schematic phase diagram given in Fig.~\ref{f1-xfig}. Except for the case 
where $\lambda_e=0$ the gaps tend toward a constant value as $M$ is increased.}
\label{fig-212121}
\end{figure}

 In Fig.~\ref{fig-212121}, we show for the case $\ell=1$ and $N=2$, the finite 
size behavior for the mass gaps as a function of $M$, for $M < 60$, and 
for some values of $\lambda_o,\lambda_e$. 
We see in this figure that in general, for $\lambda_e \neq 0$, the mass gaps tend towards a constant 
value, indicating the model is gapped. Only when
$\lambda_e=0$, we have a linear decay with $\ln M$.

In the cases where $\ell=3,5,7$ and $9$, we calculate the dynamical critical exponent $z$ from the gaps evaluated up to $M=10000$. Part of the results obtained 
from a linear fitting for the  sizes $ 3000 <M<10000$, are shown in Table 1. 
 These calculations can be extended to even larger lattice sizes 
by using the numerical method introduced in \cite{powerfull1}.
\begin{table}
  \begin{tabular}{|l|l|l|l|l|l|l|}
    \hline
    \multirow{2}{*}{$\lambda_o,\lambda_e$} &
      \multicolumn{3}{c|}{$N=2$} &
      \multicolumn{3}{c|}{$N=3$}   \\
    & $\ell=3$ & $\ell=5$ & $\ell=7$ & $\ell=3$ & $\ell=5$ & $\ell=7$ \\
    \hline
 $\lambda_o=0,\lambda_e=1$  & 1.497 & 1.994 & 2.485 & 0.998 & 1.329 & 1.669 \\
   & (3/2) &  (2) & (5/2) & (1) & (4/3) & (5/3) \\
    \hline
 $\lambda_o=1/100,\lambda_e=1$  & 1.001 & 1.513 & 2.014 & 0.666 & 1.001 & 1.335 \\
   & (1) &  (3/2) & (2) & (2/3) & (1) & (4/3) \\
    \hline
 $\lambda_o=1,\lambda_e=1$  & 1.000 & 1.501 & 2.004 & 0.666 & 0.998 & 1.330 \\
   & (1) &  (3/2) & (2) & (2/3) & (1) & (4/3) \\
    \hline
 $\lambda_o=1,\lambda_e=1/100$  & 0.999 & 1.456 & 1.992 & 0.666 & 0.995 & 1.328 \\
   & (1) &  (3/2) & (2) & (2/3) & (1) & (4/3) \\
    \hline
 $\lambda_o=1,\lambda_e=0$  & 0.999 & 1.496 & 1.994 & 0.999 & 1.496 & 1.994 \\
   & (1) &  (3/2) & (2) & (1) & (3/2) & (2) \\
    \hline
  \end{tabular}
\caption{
The dynamical critical exponent $z$ evaluated for the Hamiltonian \rf{3.11}, 
for some values of $\lambda_o$,$\lambda_e$. The results are  for the $Z(2)$ and $Z(3)$ models. 
We also show in the table (in parenthesis) the conjectured values (see text).}
\label{table}
\end{table}
The results in this table are for the fermionic case $N=2$ and for $Z(3)$ parafermionic models. 

In summary the results for $\ell$ odd \rf{3.18}, although derived numerically are similar with the exact results obtained in the case where $\ell$ is even
 \rf{3.53a}, and the phase diagram is the one shown schematically in
Fig.~\ref{f1-xfig}.

\subsection{Other models}

Let us consider models with range of interactions 
$\{r_i\}=(2,1,2,2,1,2,2,\ldots)$. A possible interesting realization for 
the fermionic case $N=2$, is given by the generators
\be \label{3.19}
h_{3i-2}^{(2)}=h_y\sigma_i^y,
h_{3i-1}^{(1)}=h_x\sigma_i^x,
h_{3i}^{(2)}=J_z\sigma_i^z\sigma_{i+1}^z,
\ee
The model for ($M+1$) sites is the quantum chain
\bea \label{3.20}
&&H = -\sum_{i=1}^M h_i^{(r_i)} = \nonumber \\
&&-h_y\sum_{i=1}^{M/3} \sigma_i^y 
-h_x\sum_{i=1}^{M/3} \sigma_i^x 
-J_z\sum_{i=1}^{M/3} \sigma_i^z \sigma_{i+1}^z, 
\ee
where $M$ is a multiple of 3. This Hamiltonian corresponds to the Ising 
quantum chain with two transverse fields ($h_x,h_y$).

The critical properties of the model is obtained from the canonical 
transformation $(\sigma^x,\sigma^y,\sigma^z) \rightarrow 
(\tilde{\sigma}^x,
\tilde{\sigma}^y,
\tilde{\sigma}^z)$,
where 
\bea \label{3.21}
&&\tilde{\sigma}_i^z = -\sigma_i^z, \quad
\tilde{\sigma}_i^x = \frac{h_x \sigma_i^x +  h_y \sigma_i^y} 
{ \sqrt{h_x^2 + h_y^2} }, \nonumber \\
&&\tilde{\sigma}_i^y = \frac{h_y\sigma_i^x +  h_x \sigma_i^y}
{\sqrt{h_x^2 + h_y^2}},
\eea
that give us 
\be \label{3.21}
H=-\sum_{i=1}^{M/3}(J_z \sigma_i^z \sigma_{i+1}^z + \tilde{h}_x \sigma_i^x), 
\quad \mbox{where} \quad 
\tilde{h}_x = \sqrt{h_x^2+h_y^2}, \nonumber
\ee
recovering the standard quantum Ising chain in a transverse field 
$\tilde{h}_x$. This model is critical when 
$J_z= \tilde{h}^x = \sqrt{h_x^2+h_y^2}$. At the isotropic point $J_z=h_x= h_y$,  the model is in a disordered gapped phase. 

For the parafermionic cases $N>2$ we fix $\lambda_{3i-2}= \lambda_A$, 
$\lambda_{3i-1} =\lambda_B$ , $\lambda_{3i}= \lambda_C$, and 
\be \label{3.23}
h_{3i-2}^{(2)} = \lambda_C Z_i^+X_i, \quad h_{3i-1}^{(1)}=\lambda_AX_i, \quad
h_{3i}^{(2)} = \lambda_B Z_iZ_{i+1}^+,\nonumber
\ee
where $Z^+= Z^{N-1}$, and $X,Z$ are the $N\times N$ matrices satisfying the $Z(N)$ 
algebra \rf{2.5}. The generators in the above equation satisfy the 
algebra \rf{2.2a} with the ranges $\{r_i\}=(1,2,2,1,2,2,\ldots)$. The 
Hamiltonian is given by 
\be \label{3.24}
H = -\sum_{i=1}^{M/3} \lambda_C Z_i^+X_i - 
\sum_{i=1}^{M/3} ( \lambda_A X_i + \lambda_BZ_iZ_{i+1}^+),
\ee
that correspond to the extension of the free parafermionic Baxter chains 
($p=1$) with couplings $\lambda_A$ and $\lambda_B$ \rf{2.7} in the presence 
of the "transverse field" $\sum_i \lambda_C Z_i^+X_i$.

Defining the transformation 
\be \label{3.25}
\tilde{h}_{2i-1}^{(1)} = \frac{\lambda_C h_{3i-2}^{(2)} + 
\lambda_A h_{3i-1}^{(1)}} {(\lambda_C^N + \lambda_A^N)^{1/N}}, 
\quad \tilde{h}_{2i}^{(1)} = h_{3i}^{(2)},
\ee
we now have
\be \label{3.26}
H= - \sum_{i=1}^{M/3}\left((\lambda_C^N + \lambda_A^N)^{1/N} \tilde{h}_{2i-1}^{(1)} + 
\lambda_B \tilde{h}_{2i}^{(1)} \right).
\ee
A direct check show us that the new generators \rf{3.25} satisfy 
$[\tilde{h}_i^{(1)}]^N =1$ and share  the same algebra with the parafermionic model 
with uniform range $p=1$ \rf{2.29}.  
The known results of the standard Baxter chain \cite{baxter1,AB1,AB2}, 
tell us that the model is critical when 
$\lambda_C^N + \lambda_A^N = \lambda_B^N$, being gapped elsewhere.

We can extend these last results for the $Z(N)$ models with range of 
interactions 
$\{r_i\} = (\ell,\ell-1,\ldots,1,\ell,\ell-1,\ldots,1,\ell,\ldots)$.
In this case the generators $h_i^{(r_i)}$ are transformed by the 
generalization of \rf{3.25}

\bea \label{3.27}
\tilde{h}_{2i-1}^{(1)} &=& 
\frac
{ \lambda_1 h_i^{(\ell)} + \lambda_2h_{i+1}^{(\ell-1)} + \cdots 
\lambda_{\ell} h_{i+\ell-1}^{(1)} } 
{(\lambda_1^N+\lambda_2^N + \cdots \lambda_{\ell}^N )^{1/N} },
 \nonumber \\
\tilde{h}_{2i}^{(1)} &=& \lambda_{\ell+1} h_{(\ell+1)i}^{(\ell)},
\ee
and the same Hamiltonian is again equivalent to an effective uniform range 
$p=1$ parafermionic model
\bea \label{3.28}
H &=& - \sum_{i=1}^{M/(\ell+1)} \left(
(\lambda_1^N+\lambda_2^N + \cdots + \lambda_{\ell}^N)^{1/N} h_{2i-1}^{(1)} 
\right. \nonumber \\
 &+& \left. \lambda_{\ell+1}h_{2i}^{(1)} 
\right), 
\eea
being critical when 
$\lambda_1^N+\lambda_2^N + \cdots +\lambda_{\ell}^N = \lambda_{\ell+1}^N$.

The results presented in this section are just examples of models with general 
 range of interaction. Some of them have a phase diagram that can be derived 
analytically, but in general a numerical analysis is necessary. This is not 
difficult since the eigenspectra is given in terms of roots of a 
polynomial $P_M(z)$, with recursion relation given in \rf{2.36}.

\section{Conclusions}

A large family of spin of $Z(N)$-invariant quantum chains with multispin interactions have a 
free-particle eigenspectra. The common ingredient of all  these models is the fact 
that they   satisfy a $Z(N)$ exchange algebra. 

In all these free-particle models the multispin coupling  
constants $\{\lambda_i\}$ are 
arbitrary for each site ($i=1,2,\ldots,M$), but the number of spins involved 
in the multispin interaction (range of interaction) is uniform, i. e., 
 is the same 
for all the lattices sites. These Hamiltonians are Hermitian and fermionic for 
$N=2$, and for $N>2$ they are parafermionic and non Hermitian.

In this paper we extend even more these families of free-particle models by 
considering now models where the range of the  multispin interactions $\{r_i\}$ 
 depends on the lattice sites ($i=1,2,\ldots,N$).

We search for the extensions of the exchange algebras that give models that still keep the integrability as well as the  free-particle spectra. In order to derive these conditions we look for the general constraints that the $Z(N)$ generators attached to the 
sites have to satisfy. Although these conditions in general was given 
in \cite{network} our derivation was done in a direct form. 

For the one  dimensional models, with site-dependent ranges $\{r_i\}$, the 
range  are restricted   to the RSOS paths where  
$r_{i+1} \geq r_i -1$ ($i=1,\ldots,M$).    This  condition enable us to 
produce nice examples, like the ones presented in  Sec.~III.1, where the  Hamiltonians 
have  a quite large global degeneracy. 

In order to exploit the physics of these models we studied in detail the 
models where all coupling constants $\lambda_o$ ($\lambda_e$) and the 
range of interactions $r_o$ ($r_e$) of the odd (even) sites are constant.
We use algebraic transformations of the generators and also numerical 
calculations to evaluate  the energy gaps of the models.  The numerical calculations were done for large lattice sizes ($M \sim 10^4$), and can be  extended to 
  even larger lattice sizes thanks to the powerful method introduced in \cite{powerfull1} (see also \cite{randonp2}) to evaluate the 
larger roots of polynomials. 

Our results show that for the ranges $r_o = \ell +1$ and $r_e=\ell$, with $\ell >1$, the 
 fermionic ($N=2$) and parafermionic ($N>2$) models are critical. They belongs to a 
universality class of critical behavior where the dynamical critical exponent is 
$z= (\ell+1)/2N$ if $\lambda_e \neq 0$ and $z= (\ell+3)/2N$ if $\lambda_e =0$. For the value 
$\ell =1$ ($r_o=2, r_e=1$), and $\lambda_e \neq 0$  the model is gapped, while when $\lambda_e=0$  the model is critical with $z=2/N$. Most probably 
whenever $z=1$ the models are conformally invariant as happens in the uniform range models \cite{multi-conf}. 

In general however the models although exact integrable are non conformally invariant at their critical points since $z\neq 1$. Since most of the studied 
critical quantum chains are conformally invariant, the  models presented in 
this paper are  useful to probe new physical ideas and also for the production of toy models 
of many-body interactions for testing numerical algorithms.


\begin{acknowledgments}
We thank
 discussions with Jos\'e A. Hoyos and Rodrigo Pimenta.
This work was supported  in part by the Brazilian agencies FAPESP (Proc.2024/05981-4) and  CNPq. 
\end{acknowledgments}

\appendix

\section{Inversion relations}

Taking into account the involution of the charges \rf{2.12}, i. e.,
$[Q_M^{(\ell)},Q_M^{(\ell')}]=0$ ($\ell,\ell'=0,1,\ldots$), we are going to obtain 
in this appendix the conditions that ensure the inverse relation \rf{2.29b}.
Instead of using graph theory as in \cite{network}, similarly as we did 
in section II, we are going to obtain these conditions directly from the 
expansions of the charges \rf{2.27}. 

We restrict here for the case $N=2$, where the inversion relation is 
\bea \label{a1}
T_M(u)&=&G_M(u)G_M(-u) \nonumber \\ 
&=& \sum_{\ell,\ell'}^{\bar{M}} 
u^{\ell+\ell'} (-)^\ell Q_M^{(\ell)}Q_M^{(\ell')}=P_M(u^2).
\eea
Since $[Q_M^{(\ell)},Q_M^{(\ell')}]=0$, we can write
\be \label{a2}
T_M(u)= \sum_{\ell=0}^{\bar{M}} \hat{Z}^{\ell} z^{\ell}, \quad z= u^2,
\ee
and $\hat{Z}^{(\ell)}$ is an operator given in terms of the conserved charges 
$\{Q_M^{(\ell)}\}$:
\be \label{a3}
\hat{Z}^{(\ell)} = \sum_{j=0}^{2\ell} (-)^j Q^{(2\ell-j)} Q_M^{(j)}, 
\quad \ell=0,1,\ldots .
\ee
$Q_M^{(\ell)}$ are formed by the products of $\ell$ commuting generators, 
consequently $\hat{Z}^{(\ell)}$ is given by the sum of all the words 
$W^{(2\ell)}$ containing $2\ell$ generators distributed in two commuting groups 
with ($2\ell-j$) and $j$ generators ($j=0,1,\ldots,2\ell$), respectively.
The inverse relation \rf{a1} holds only if $\hat{Z}^{(\ell)}$ is an scalar 
for any $\ell$.

In general the generators forming the words $W^{(2\ell)}$ can be separated 
in subgroups $g_i$ ($i=1,2,\ldots$), where all the operators inside a given 
subgroup $g_i$ commute with the others belonging to distinct subgroups 
 ($g_i \neq g_j$):
\be \label{a4}
W^{(2\ell)} = g_1 g_2 \cdots  .
\ee
\begin{figure} [htb]
\centering
\includegraphics[width=0.45\textwidth]{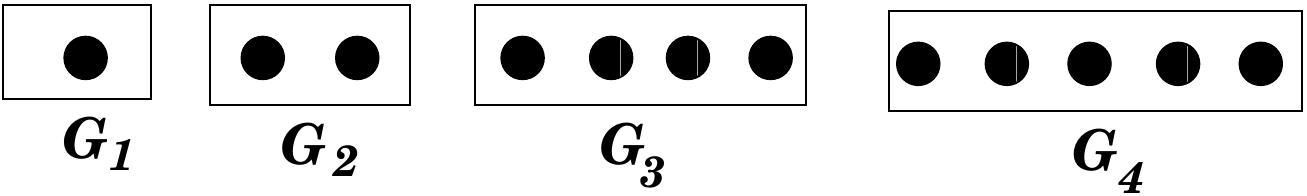} 
\caption{
The word $W^{(12)}= h_1h_2\cdots h_{12}$ in \rf{a4} formed by the subgroup 
$g_1=h_1$, $g_2=h_2h_3$, $g_3=h_4h_5h_6h_7$ and $g_4=h_8h_9h_{10}h_{11}h_{12}$.
 The generators inside a given subgroup commute with the ones in 
distinct subgroups.} \label{I1}
\end{figure}

As an example we show in Fig.~\ref{I1} the word $W^{(12)}=h_1h_2\cdots h_{12}$ 
formed by the subgroups $g_1=h_1$, $g_2=h_2h_3$, $g_3=h_4h_5h_6h_7$, and 
$g_4=h_8h_9h_{10}h_{11}h_{12}$.  
The order we write the subgroups in \rf{a4} does not 
matter due to the commutation of the generators in distinct subgroups.

Let us consider separately the possible subgroups that may appear in the 
word $W^{(2\ell)}$ given in \rf{a4}. We want to see what are the conditions 
on the algebra of $\{h_i\}$ that ensures that the words $W^{(2\ell)}$ 
are just c-numbers.

{\bf a)}  Subgroup $g_i$ containing a single generator $h_m$ (like $g_1$ in 
Fig.~\ref{I1}). Words contained in this subgroup do not appear in 
$\hat{Z}^{(\ell)}$, since keeping all the others generators, $h_m$ will appear 
in the  expansions of $G_M(u)$ and $G_M(-u)$ and the addition of both 
contributions vanishes.

{\bf b)} Subgroup with two generators $h_m,h_n$, with $\{h_m,h_n\}=0$, like 
in $g_2$ in Fig.~\ref{I1}. In this case the same word will appear in 
$\hat{Z}^{(\ell)}$ twice, one where the order of the two operators is 
$h_mh_n$ and he other $h_nh_m$. The total contribution is $h_mh_n+h_nh_m=0$. 

\begin{figure} [htb]
\centering
\includegraphics[width=0.45\textwidth]{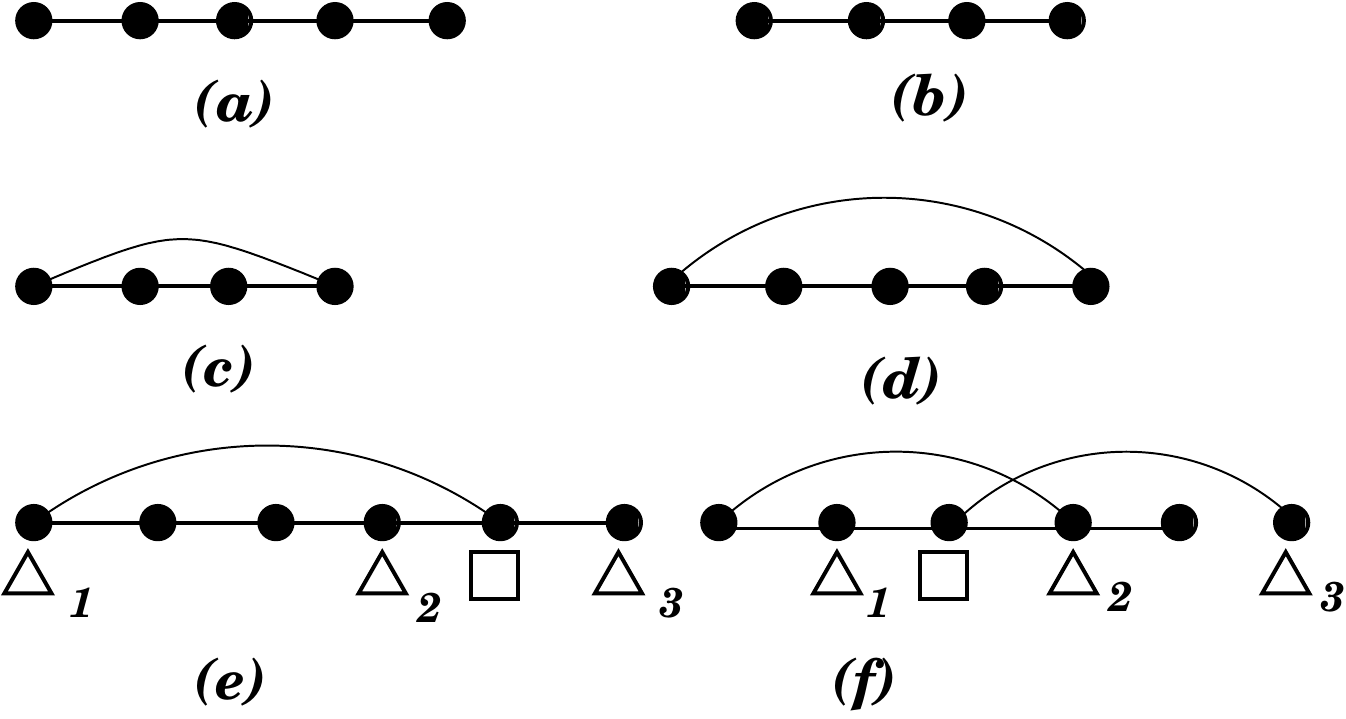} 
\caption{
Subgroups $g= h_1h_2 \cdots h_m$. In (a) the number of generators is odd and 
in  (b) 
it is even. In (c) and (d) we have subgroups forming 
closed loops of non-commuting operators. In (e) and (f) we have non allowed 
subgroups, since they violate the conditions for integrability (see text).}
\label{I2}
\end{figure}

{\bf c)}  Subgroup with an odd number of generators as in Fig.~\ref{I2}d. 
Since it is not possible to split the generators in two commuting subsets, 
this subgroup never appears.

{\bf d)} Subgroup with an even number of $m$ generators, as in Fig.~\ref{I2}b, 
$g=h_1h_2\cdot h_m$. We can separate $g$ in two commuting sets 
$g=g_og_{e}=g_{e}g_o$ where $g_o=h_1h_3\cdots h_{m/2}$ and 
$g_{e}=h_2h_4\cdots h_m$.  
It is simple to see that $g_og_{e} = -g_{e}g_o$. This imply that the 
additions of the words where the subgroups $g_o$ and $g_{e}$ appear in 
reverse order vanishes.

{\bf e)} Subgroup $g=h_1h_2\cdots h_m$ where the links connecting the 
non-commuting generators form a closed loop, as in Fig.~\ref{I2}c and 
Fig.~\ref{I2}d. 
In the case where $m$ is even, as in Fig.~\ref{I2}c,  we can separate
$g=g_og_{e}=g_eg_o$ where $g_o=h_1h_3\cdots h_{m/2}$ and
$g_{e}=h_2h_4\cdots h_m$. However distinct from the case b), now 
$g_og_e=g_eg_o$ and the addition of the words with the two possible ordering 
give us $g_eg_o+g_og_e \neq 0$. This means we should not have in the algebra 
non-commutative loops of operators like in Fig.~\ref{I2}c or Fig.~\ref{I2}d 
in order to satisfy the inversion relation. This type of loops happens 
when the generators are attached to the sites of periodic lattices. This 
imply that the inversion relations, at least defined as in \rf{2.30}, is 
not valid in this case. More general inversion relations, distinct from 
\rf{2.30} is expected in this case.

{\bf f)} Subgroup $g$ containing closed loops of non-commuting operators 
like Fig.~\ref{I2}e and Fig.~\ref{I2}f. This type of subgroup does not exist 
since in this case we have a generator ($\square $) connected with 3 
commuting generators ($\triangle_1,\triangle_2$ and $\triangle_3$).
  This  violates the condition derived 
in section II, that ensures the involution 
$[Q_M^{(\ell)},Q_M^{(\ell')}]=0$ of the charges.

{\bf g)} Finally the remaining possibilities to obtain a non-zero contribution 
in $\hat{Z}^{(\ell)}$ are given by the products
\be \label{a5}
h_{i_1}^2 h_{i_2}^2 \cdots h_{i_{\ell}}^2 = \lambda_{i_1}^2 \lambda_{i_2}^2 
\cdots \lambda_{i_{\ell}}^2,
\ee
which are c-numbers due to the closure relation 
$h_i^2=\lambda_i^2$ ($i=1,2,\ldots$). 
These terms come from the product $(-)^{\ell} \; Q_M^{(\ell)} 
Q_M^{(\ell)}= (-)^{\ell}[Q_M^{(\ell)}]^2$. 
The square of $Q_M^{(\ell)}$ cancels all the contributions coming from the 
anti commuting terms in $Q_M^{(\ell)}$ and we only have the 
commuting ones, i. e.,
\be \label{a6} 
\hat{Z}^{(\ell)} = Z^{(\ell)} = (-)^{\ell} 
\sum_{\{i_1,i_2,\ldots,i_{\ell}\}}^{(*)} 
\lambda_{i_1}^2 \lambda_{i_2}^2 \cdots \lambda_{i_{\ell}}^2,
\ee
where $(*)$ denotes the sum over all independent possibilities of products 
of $\ell$-couplings associated to $\ell$ commuting  generators 
$h_{i_1},h_{i_2}, \ldots,h_{i_{\ell}}$ in $Q_M^{(\ell)}$. 
We can then write, from \rf{a1}-\rf{a3}, $G_M(u)$ as a polynomial in $u^2$
\be \label{a7}
T_M(u) = G_M(u)G_M(-u) = \sum_{\ell=0}^{\bar{M}} 
{\hat{Z}}^{\ell} u^{\ell} = P_M(u^2).
\ee

 In the symmetric case $\lambda_1=\lambda_2=\cdots =\lambda_M =1$, 
$Z^{\ell}= d(\ell,M)$ is the number of $\ell$-products in the $Q_M^{(\ell)}$ 
charge of the  $M$-generators algebra. 

In summary, if the generators do not form non-commuting loops in the 
product $Q_M^{(\ell)}\dot Q_M^{(\ell')}$ and no operator do not commute with 
three commuting operators the inversion relation \rf{a1} is valid.

\end{document}